\documentclass[12pt]{spieman}  

\usepackage{graphicx}
\usepackage{subcaption}
\usepackage{newtxtext,newtxmath}
\usepackage{pifont}
\usepackage{gensymb}
\usepackage{setspace}
\usepackage{lineno}
\usepackage{tocloft}

\newcommand{\g}{\mathrm{g}}
\newcommand{\cm}{\mathrm{cm}}

\newcommand{\pc}{\mathrm{pc}}

\newcommand{\redshift}{z_\text{redshift}}

\newcommand{\pFIR}{p_{0, \text{FIR}}}

\newcommand{\ds}{\text{ds}}%
\newcommand{\ndust}{n_\text{dust}}%
\newcommand{\ndusti}{n_\text{i, dust}}%
\usepackage[numbers]{natbib}
\usepackage{graphicx}
\usepackage{setspace}
\usepackage{tocloft}
\usepackage{caption}
\title{PRIMA Vista: far-infrared polarimetry to unveil small-scale magnetohydrodynamics in extragalactic observations}

\author[a,b,*]{Diego Maglione}
\author[b]{Sergio Martin-Alvarez}
\author[c,b]{Enrique Lopez-Rodriguez}
\author[a,b]{Susan E. Clark}
\author[a,b]{Kaitlyn Karpovich}
\affil[a]{Stanford University, Physics Department, 452 Lomita Mall, Stanford, United States, 94305}
\affil[b]{Kavli Institute for Particle Astrophysics \& Cosmology (KIPAC), Stanford University, Stanford, CA 94305, USA}
\affil[c]{Department of Physics \& Astronomy, University of South Carolina, Columbia, SC 29208, USA}

\cftpagenumbersoff{figure}
\cftpagenumbersoff{table} 
\begin{document} 
\maketitle

\begin{abstract}

Magnetic fields are one of the fundamental components of the interstellar medium (ISM), and remain a challenge for building a comprehensive understanding of galaxies and their properties across cosmic time. Their synoptic study requires far-infrared polarimetric observations, which provide an unrivaled probe of the dynamics, magnetization, and structure of the coldest and densest interstellar gas and dust at small scales in galaxies---where the mass and star formation reside. We employ high-resolution cosmological magnetohydrodynamical simulations of a face-on Milky Way-like galaxy and show that the alignment of magnetic fields with ISM structures and the turbulence at 100~pc scales decrease with increasing magnetization. We make predictions for extragalactic observations by the proposed PRIMA telescope, comparing them with SOFIA observations similar to those of the SALSA survey. PRIMA will be able to better measure magnetic alignment trends previously inaccessible by SOFIA observations. We find PRIMA observations will better sample magnetic turbulence, especially in dense environments, and will be able to measure the unresolved intrinsic magnetic field orientations to $\sim6^\circ$ precision. Additionally, PRIMA will be capable of resolving observables such as the polarized fraction or the magnetic alignment down to scales comparable to the resolution of our simulations ($\sim$10~pc) for galaxies $\le0.5$ Mpc. The intrinsic polarization–dispersion relation shows PRIMA observations will suffer from significantly reduced beam depolarization. Furthermore, PRIMA will recover the correlation between increasing magnetic alignment paramater and local polarization fraction. Overall, observations of local galaxies with PRIMA will better characterize interstellar magnetic properties and constrain ISM and galaxy models, advancing our understanding of magnetism in the Universe.

\end{abstract}


{\noindent \footnotesize\textbf{*}Diego Maglione,  \linkable{dmag@stanford.edu} }

\begin{spacing}{1}   

\section{Introduction} \label{sec:intro}

Magnetic fields are critical players in the structural and dynamical properties of a galaxy's interstellar medium (ISM). Magnetic fields modify the distribution of gas across the ISM phases \citep{Iffrig2017, Ji2018}, reshape turbulent properties \citep{Evirgen2019, Federrath2012}, and have an important series of additional effects in different density regimes. In diffuse gas, magnetic fields affect the ISM porosity by suppressing the development of low-density disk cavities \citep{Kortgen2019, Martin-Alvarez2020}, modify gas fragmentation into denser structures \citep{Inoue2019}, and direct gas flows in the formation of molecular clouds \citep{Tahani2022a}. One of the most important effects of magnetic fields occurs in dense regions, where magnetic fields regulate and modify star formation rates and properties \citep{McKeeOstriker2007, Federrath2012}.

Due to the magnetic field's widespread and diverse effects across different gas phases, a multiwavelength sampling of magnetic fields across the multi-phase ISM is required to build a comprehensive picture of their physical role. The radio and far infrared (FIR) regimes are notably complementary, with each serving as a probe of magnetic fields associated with star formation at different length scales and ISM phases \citep{Martin-Alvarez2024}. Radio observations preferentially trace warm and diffuse gas, following star formation driven by supernova-accelerated cosmic ray electrons, and are subject to caveats related to their spectra and distribution \citep{Dacunha2024}. Instead, FIR observations are dominated by thermal dust emission, which is weighted toward cold and dense regions of the ISM where most of the mass and star formation reside. 

In the past few years, unprecedented FIR ($53-250\,\mu$m) polarimetric observations of galaxies \citep{Lopez-Rodriguez2022b}  by the High-resolution Airborne Wideband Camera-Plus (HAWC+) instrument on the Stratospheric Observatory For Infrared Astronomy (SOFIA), in combination with radio polarimetric observations, showed the potential of such complementarity in studying turbulence, dynamics, star formation, and magnetic field properties. The Survey of extragALactic magnetiSm with SOFIA (SALSA) showed that the FIR B-field orientations are cospatial with the gas density distribution in the spiral arms and star formation regions of galaxies \citep{Lopez-Rodriguez2022b, Borlaff2023}. In addition, SALSA found that FIR B-field magnetic alignment parameters, which trace the local degree of coherence of the magnetic field orientation with galactic-scale morphological structures, are significantly lower than those traced at radio wavelengths \citep{Borlaff2023}. This result indicates that the FIR polarimetric observations are more sensitive to the turbulence and star formation activity within the beam ($\sim300$ pc) of the SOFIA observations of nearby galaxies.

However, given the spatial resolution \citep[$\sim300$ pc; typical resolution for the SALSA sample;][]{Lopez-Rodriguez2022b} of the observations, both star formation regions and the turbulent cascade remain unresolved. Turbulence plays an important role in structuring the magnetic field down to significantly smaller scales \citep{Han_2004}. This turbulence, injected by supernova feedback at $\sim$$100\,\pc$ scales \citep{Martizzi2016}, is transferred to magnetic fluctuations \citep{Martin-Alvarez2018, Brandenburg2023}. Resolving even the upper end of this cascade in external galaxies therefore requires higher-resolution observations.

Synthetic polarized emission from magnetically aligned non-spherical dust grains using Auriga's magnetohydrodynamical (MHD) simulations by \cite{Vandenbroucke2021} demonstrated that FIR polarization primarily traces ordered magnetic fields in low-density regions of Milky-Way-like galaxies at scales of approximately $\sim$1 kpc. Their results highlight how the plane-of-sky magnetic field orientation remains coherent primarily in low-density environments, suggesting reduced polarization signals in denser regions due to increased magnetic field tangling. As the turbulence scale is well-below these spatial resolution, further high-resolution simulations are needed. Thus, FIR polarimetric observations with spatial resolutions $<50$\,pc are required to understand the physical mechanisms driving the B-field angular dispersions across the density structures of a galaxy's disk.

Due to the recent decommissioning of SOFIA, FIR ($53-250\,\mu$m) polarimetric observations have become inaccessible. Although ground-based FIR/sub-mm ($>450\,\mu$m) observatories have measured the B-fields in several nearby galaxies \citep{Pattle2021,Clements2024}, only a handful of galaxies are accessible due to their low sensitivity. While some ground-based instruments (e.g., JCMT at $850\,\mu$m with $13''$ resolution) offer competitive spatial resolutions when compared to SOFIA’s bands, the overall effective resolution for ground-based FIR/sub-mm observations remains challenging due to lower sensitivity. The proposed PRobe far Infrared Mission for Astrophysics; \citep[PRIMA;][]{Moullet2023} mission will build on SOFIA’s legacy by delivering unprecedented sensitivity and effective resolution studies of magnetic fields across diverse galactic environments. It is in this context that numerical models are useful for interpreting current and future FIR observations. They can aid in understanding the imprint of local structures such as the Local Bubble on galactic polarized dust emission \citep{Maconi2023}, or explain the degree of polarization of gas clouds \citep{Juvela2019}. For galactic and extragalactic observations, synthetic observables such as the dust polarization fraction also exhibit important correlations with the gas column and FIR intensities, varying with different galaxy inclinations \citep{Vandenbroucke2021}.

\begin{figure}
\centering
\includegraphics[width = \columnwidth]{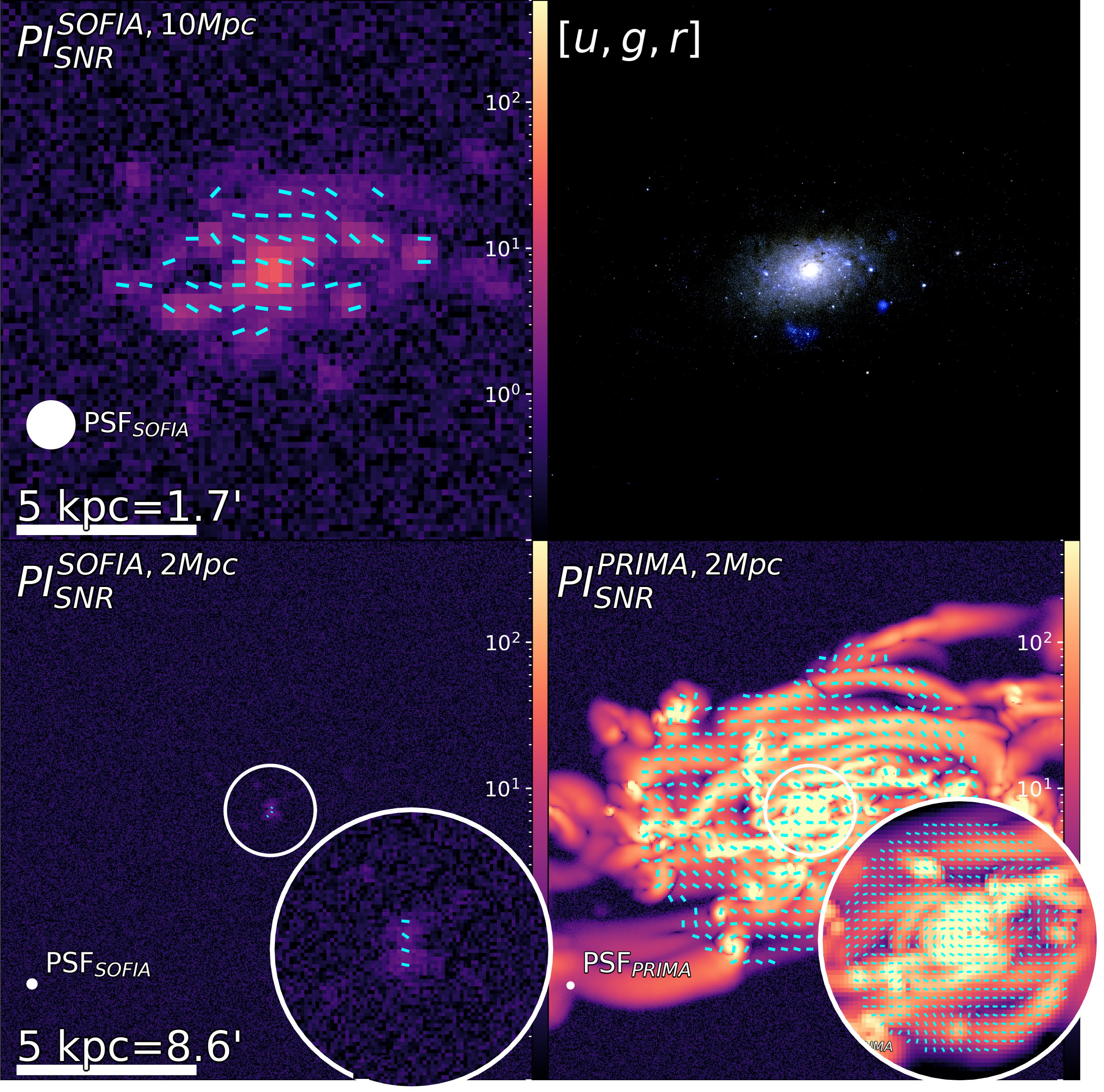}
\caption[Observational Comparison of the Polarized Intensity Signal-to-Noise Between SOFIA and PRIMA]{Observational comparison of the polarized intensity (PI) signal-to-noise (SNR) between SOFIA and PRIMA for observations of extragalactic sources through synthetic observations of the simulated MB12 galaxy model. ({Top left}) SOFIA-like observation at 10~Mpc, with low resolution and higher SNR of unresolved structures. ({Bottom left}) SOFIA-like observation at 2~Mpc, with higher spatial resolution but lower SNR due to sensitivity constraints. Inset panel zooms into the central region of the image, as highlighted by the central white circle. ({Bottom right}) PRIMA-like at observation 2~Mpc, with higher sensitivity leading to higher effective spatial resolutions. Inset panel zooms into the central region of the image, as highlighted by the central white circle. Colorscale values are displaced downwards by exactly 1 dex to reduce saturation. ({Top right}) Optical synthetic observation of the galaxy in the [u,g,r] SDSS filters. Across al PI SNR panels, overlaid quivers show the orientation of the magnetic field as inferred from the polarized intensity. Due to the high effective resolution of PRIMA, we downsample the quivers in the main bottom right panel by a factor of 5.}
\label{fig:front_figure}
\end{figure}

In this work, we employ a suite of various cosmological MHD simulations of a face-on, nearby ($<10$ Mpc) Milky Way-like galaxy to examine various observables accessible by FIR polarimetry. We focus on magnetic turbulence, the relative alignment of magnetic and density structures in the ISM, observational estimates of magnetic disorder, and the effect of gas density and magnetization strength on these quantities. We compare our predictions for the PRIMA probe with those of SOFIA and with the ground truth of our simulations to show the unprecedented potential of the proposed PRIMA mission. A summary example of the MHD simulations and mock observations is shown in Fig. \ref{fig:front_figure}. The manuscript is structured as follows: we describe the numerical methods and the setup of our cosmological MHD simulations in Section~\ref{sec:methods}, detailing the varying magnetization of our galaxy models.
In Section~\ref{sec:results}, we present our main results, studying magnetic field alignment with surface density structures and turbulence across density environments. Finally, Section~\ref{sec:conclusion} summarizes our findings and discusses their implications for understanding galactic magnetism.

\section{Methods} \label{sec:methods}
\subsection{The RAMSES Code}

In this study, we employ previously introduced numerical MHD simulations \citep{Martin-Alvarez2020, Martin-Alvarez2021}. We limit our analysis to galaxies in their late evolutionary stages, corresponding to the period after the main galaxy of interest has concluded its main formation. From this point onwards, the galaxies only secularly evolve without further significant growth or merger events. Our simulations are generated using {\sc ramses}, which treats the stellar particles and dark matter particles as collisionless elements, whereas baryonic gas is treated through an Eulerian grid. These different components have their evolution coupled through the gravity solver. The {\sc ramses} MHD grid incorporates adaptive mesh refinement (AMR) into its octree structure, allowing for higher resolution in regions of interest. The finite discretization of the simulated grid results in non-zero numerical resistivity in our MHD solver \citep{Teyssier2006}. With the physical magnetic diffusivity set to $\eta = 0$, any remaining magnetic diffusion effects in our simulation are numerical artifacts stemming from this discretization resistivity. This MHD solver uses the constrained transport method \citep{Teyssier2006, Fromang2006} to model the magnetic field, which satisfies the $\vec{\nabla} \cdot \vec{B} = 0$ constraint for ideal MHD down to numerical precision \citep{Martin-Alvarez2018, Martin-Alvarez2020}. This constraint must be satisfied to avoid spurious magnetic monopoles and to ensure that the method preserves conserved quantities while avoiding other numerical artifacts \citep{Toth2000}.

\subsection{Cosmological Initial Conditions and Simulated Domain}

The MHD simulations we study employ the NUT galaxy initial conditions \citep[ICs;][]{Powell2011}: a spiral galaxy similar in properties to the Milky Way, the magnetic evolution and characteristics of which have been studied extensively in \cite[e.g.,][]{Martin-Alvarez2018,Martin-Alvarez2020}. The ICs correspond to a cubic box with  12.5 comoving Mpc (cMpc) on each side. The NUT galaxy forms in the center of a spherical zoom region with a diameter of $4.5$~cMpc. The adaptive refinement octree grid in our simulations is allowed to resolve a minimum physical cell size of approximately $\sim$$10$~pc. This level of refinement demonstrates turbulent characteristics comparable to galactic models using uniform grid refinement \citep{Martin-Alvarez2022}, although we note that convergence may still require higher resolution by at least a few dex \citep{Kortgen2017, Gent2023}.

\subsection{Different Magnetization Models for the Magnetic Field in the Galaxy}
For our study, we employ five different models of the formation of the galaxy, each featuring different magnetizations implemented by varying the comoving strength of the initial magnetic field ($B_0$), uniformly permeating the simulation volume. Our models are labelled: MB10, with \(B_0\sim10^{-10}\,\)G; MB11, with \(B_0\sim10^{-11}\,\)G; MB12, with \(B_0\sim10^{-12}\,\)G; MB20, with \(B_0\sim10^{-20}\,\)G; and MBinj, with \(B_0\sim10^{-20}\,\)G and astrophysical magnetization seeded by supernova (SN) remnants. In the MBinj model, each SN injects 1\% of their energy into their surrounding cells \citep[see Appendix A of][]{Martin-Alvarez2021} to reproduce the approximate magnetic field strength of SN remnants  \citep[$\sim$$10^{-5}$ G; ][]{Parizot2006}. MB20 is a galactic model of negligible magnetization, while the MB10 model serves as a useful representation of galaxies under extreme magnetization levels. MB10 is still allowed by current magnetization constraints \citep{PlanckCollaboration2015}, but leads to an unrealistic reionization history of the universe, primordial matter clustering, and severe variations of galactic properties \citep{Martin-Alvarez2020, KMA2021, Sanati2024}.

\subsection{Generation of Far-Infrared Synthetic Observations} \label{ss:mocks}
As a proxy for FIR polarimetric emission, we employ a geometric approximation to retrieve the polarized dust emission column \citep{LD1985,FP2000,PlanckXX2015,Chen2016,King2018,Lopez-Rodriguez2020,Martin-Alvarez2024}. For each cell ($i$) of the simulated domain, we assume a contribution to each of the Stokes parameters $I$, $Q$, and $U$:
\begin{equation}
    I_{i,\rm{FIR}} =  \ndusti \left[1.0 - \pFIR \left(\frac{{B_{i, x}^2 + B_{i, y}^2}}{{B_i^2}} - \frac{2}{3}\right)\right] \ds_i,
    \label{eq:ImockFIRdust}
\end{equation}
\begin{equation}
    Q_{i,\rm{FIR}} = \ndusti\ \pFIR \ \left(\frac{{B_{i, y}^2 - B_{i, x}^2}}{{B_i^2}}\right) \ds_i,
    \label{eq:QmockFIRdust}
\end{equation}
\begin{equation}
   U_{i,\rm{FIR}} = \ndusti\ \pFIR\  \left( \frac{{2B_{i, x} B_{i, y}}}{{B_i^2}} \right) \ds_i.
    \label{eq:UmockFIRdust}
\end{equation}

\noindent Here, each cell ($i$) features its own value of dust number density ($\ndust$), cell size along the line of sight ($\ds$), and magnetic field strength ($B$), which is further decomposed for each observation into its components perpendicular to the line of sight (LOS), $B_x$ and $B_y$. In our face-on geometry, we define the $z$-axis as perpendicular to the galactic plane, so that $ds\equiv dz$ along the line of sight. We set the maximum polarization fraction to $\pFIR = 0.25$, following the maximum polarized dust emission fraction measured in \textit{Planck} dust polarimetric observations of the Milky Way \citep{PlanckXII2020}. We note that the polarization fraction does not affect the polarization angle used to determine the magnetic field orientation, as this angle depends on the ratio between the Stokes parameters $Q$ and $U$. 

For each cell, the dust number density is estimated following
\begin{equation}
    \ndust = \left(\rho_\text{gas}\, Z\, \eta_\text{D/M}\right) m_\text{dust}^{-1} \, f_\text{cut} (T),
    \label{eq:dust}
\end{equation}
\noindent where $\rho_\text{gas}$ is the cell gas density, is $Z$ its metallicity, and $T$ is its gas temperature. We assume a constant dust-to-metal ratio of $\eta_\text{D/M}=0.4$ \citep{Dwek1998, Draine2007b}, with dust grain mass $m_\text{dust} = 1.26\times10^{-14}\,\g$ \citep[obtained by assuming radius $0.1\,\mu m$ and density $3\,\g\,\cm^{-3}$;][]{Zubko2004}. $f_\text{cut} (T)$ is the function used to select the ISM phase given a gas temperature. The exploration of various functional forms for $f_\text{cut} (T)$ results in only minor variations in the FIR emission maps \citep{Martin-Alvarez2024}. We employ an ionization-dependent metal-to-dust ratio \citep{Laursen2009};  \citep[Model 3 in Appendix A,][]{Martin-Alvarez2024}. 

To generate the synthetic maps of the Stokes $I$, $Q$, and $U$ in their base projection, we compute two-dimensional face-on projected maps along the LOS by integrating the proportional volume contribution from each cell according to the volume fraction intersected by the pixel LOS. Note that these maps do not account for observational effects such as noise or resolution (Section \ref{sec:replicatingobservations}). Our calculations assume a constant distribution at galactic scales for both dust alignment with the magnetic field as well as for other intrinsic dust properties (temperature, composition, etc.); and that depolarization of the dust distribution is dominated by geometric effects and decoherence of magnetic field orientation within the studied pixel LOS and beam size, when employed. A more sophisticated treatment of dust formation and evolution in simulations \citep[e.g.,][]{Dubois2024} and their synthetic observation is beyond the scope of this paper.

The mock dust intensity (Eqs.~\ref{eq:ImockFIRdust}--\ref{eq:UmockFIRdust}) is primarily modulated by the local gas density and metallicity (Eq.~\ref{eq:dust}). The metallicity field varies both at small scales---with ISM structures such as pre-star forming clouds or enriched SN remnants featuring different values---and at galactic scales, due to metallicity gradients. The intensities from Eqs.~~\ref{eq:ImockFIRdust}--\ref{eq:UmockFIRdust} also depend on the geometry of the magnetic field in each cell -- specifically, on the magnetic field orientation with respect to the line of sight. As a result, important differences arise between the gas surface density (dependent only on the gas number density, $n_\text{gas}$) and our integrated FIR observables (also dependent on $\vec{B} / B$ and $Z$). This decorrelation is shown by Figs.~3 and 4 in \cite{Martin-Alvarez2024}. We have also explored multiple gas temperature and ionization modulations ($f(T)$ term; see \cite{Martin-Alvarez2024}). However, these have only a secondary impact on our observables, as they primarily modify regions of particularly low gas densities.


In this work, we are primarily interested in the scaling of various quantities with FIR emission, and the main inferred quantity we extract from the mock observations is the orientation of the magnetic field associated with the mock FIR polarized emission. We express the Stokes parameters in units of dust surface density ($\cm^{-2}$). This allows for a straightforward comparison with hydrogen surface density ($\cm^{-2}$). We also investigate line-of-sight depolarization effects, employing the polarization fraction.

\subsection{Turbulence Fractions} \label{sec:turbulencefractions}
To investigate turbulence in our simulated galaxies, we estimate the fraction of the gas velocity field and magnetic field that is attributable to turbulent fluctuations. We compute turbulent fractions employing a fixed spatial scale of $\mathcal{L}=100$~pc \citep{Martin-Alvarez2022}. For each cell in the simulation, and for each field quantity $A_{x,y,z}$, we compute its deviation from its average value in a sphere of radius $0.5\mathcal{L}$, resulting in $A_\textrm{x,y,z, turb}$. This computation employs the full 3D reconstruction of the entire AMR grid, extracting the full fraction of the simulated domain required for the measurement.

To compute the turbulent velocity and turbulent magnetic field fraction for a given cell, we compute the Cartesian turbulent components $A_\textrm{x, turb}$, $A_\textrm{y, turb}$, and $A_\textrm{z, turb}$, and obtain the turbulent fraction, $f_\textrm{turb}$, as
\begin{equation}
    f_\textrm{turb} = \frac{\sqrt{A_\textrm{x, turb}^2 + A_\textrm{y, turb}^2 + A_\textrm{z, turb}^2}}{A},
\label{eq:turb_fraction}
\end{equation}
\noindent where $A$ is the modulus of the total field, $A = \sqrt{A_\textrm{x}^2 + A_\textrm{y}^2 + A_\textrm{z}^2}$. The resulting quantities are scalar values accessible for each cell of the entire computational domain, and tractable analogously to other quantities directly followed in the simulation.

\subsection{Projected Physical Quantities} \label{sec:ext_quantities}
From each of the approximately $\sim$$15$ snapshots available for the MB10, MB11, MB12, MB20, and MBinj models after redshift $\redshift = 2$, which are sampled approximately every $\sim$$150$ Myr, we compute two-dimensional face-on projections of various physical quantities from the studied spiral galaxy. Each projected map spans 15 kpc per side. The computed quantities are the gas surface density, the magnetic field orientations in the plane of the sky (separately for the x and y components), turbulent gas velocity fraction, turbulent magnetic field fraction, mock FIR total intensity, mock FIR Stokes $Q$, and the mock FIR Stokes $U$. Extensive quantities are integrated along the LOS, whereas the intrinsic magnetic field components in the plane of the sky ($\mathbf{B}_{\text{POS}}$) and the turbulent fractions are computed as gas density-weighted averages along the LOS. 
These quantities are extracted from the simulation employing a full reconstruction of the AMR grid, and projected into a 2D screen through a flat projection that resembles extragalactic observations. We compute the contribution of each resolution element to each pixel of the image array, accounting for the fraction of the voxel resolution element contained within each pixel. This ensures mass conservation and high accuracy down to the smallest scales resolved by the simulation.

\subsection{PRIMA and SOFIA Mock Observations} \label{sec:replicatingobservations}
\begin{figure}
\centering
\includegraphics[width = \columnwidth]{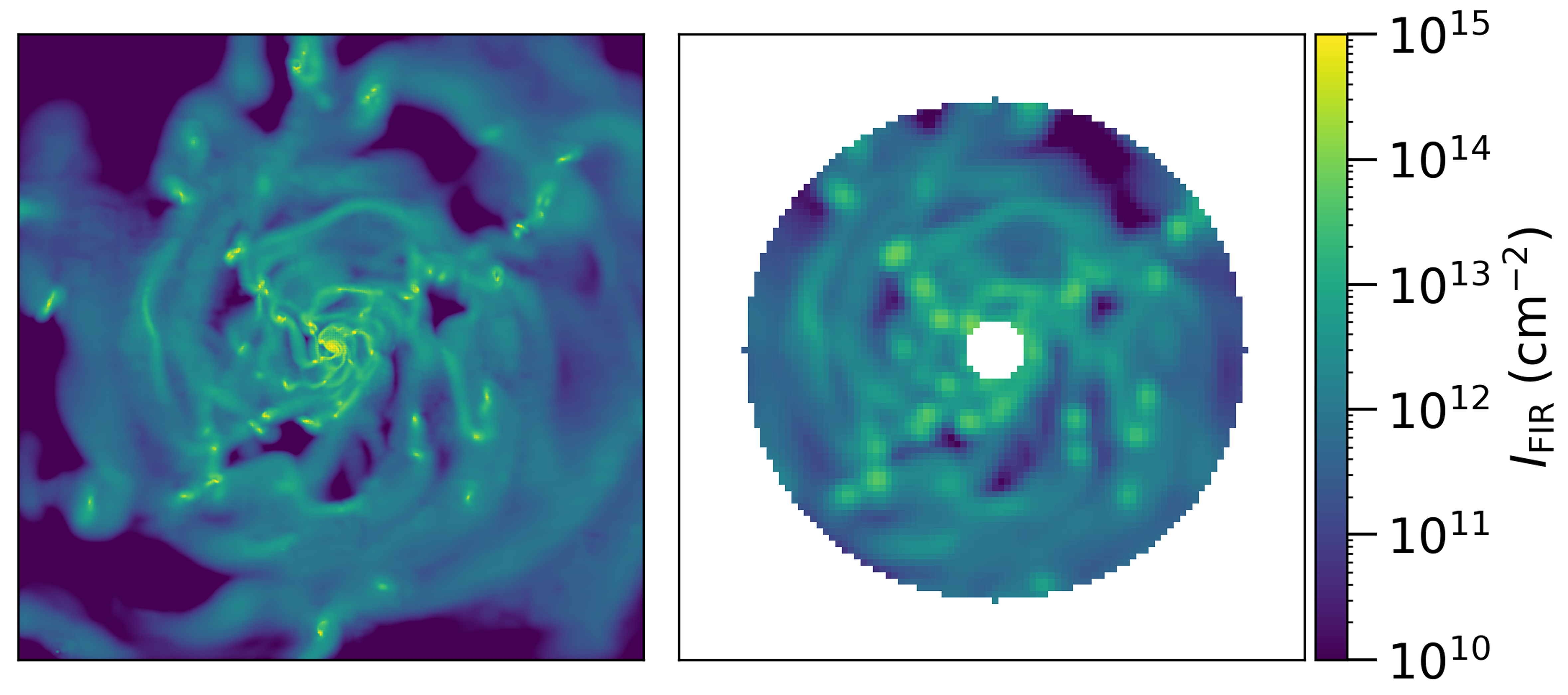}
\caption[Replicating Resolution and Applying Annulus Cuts]{Simulated galaxy and SOFIA mock observation. 
Left: $15 \times 15$~kpc$^{2}$ face-on image of the mock FIR total intensity of the NUT galaxy for the MB11 magnetization model, before mock SOFIA-like image processing. 
Right: $15 \times 15$~kpc$^{2}$ face-on image of the mock FIR total intensity of the NUT galaxy generated for the same MB11 magnetization model, after SOFIA-like image processing, which includes Gaussian smoothing and Nyquist sampling (see text for details). Both panels follow the same displayed colorbar.}
 \label{fig:replicatingsofia}
\end{figure}

To compute mock polarimetric observations, we perform Gaussian smoothing on each of the Stokes parameters as well as on the physical quantities computed. For  HAWC+/SOFIA, we assume a physical resolution of $300$~pc with a point spread function width of $13.6"$ at 154~$\mu$m \citep[capturing the approximate diffraction limit of the HAWC+/SOFIA point spread function;][]{Harper2018} at a distance of $10$~Mpc, as the representative physical scale from the SALSA Legacy program \citep{Lopez-Rodriguez2022b}. To avoid oversampling of the resulting information, we resample the images to match the resolution of the SALSA observations \citep[SOFIA Legacy Program;][]{Lopez-Rodriguez2022a}, assuming a standard Nyquist sampling of the beam \citep{Martin-Alvarez2024}. Lastly, we mask each image into an annulus with an inner radius of a physical size of 750~pc and an outer radius of 6000~pc. The annulus masking excludes from our analysis the dense galactic centers, which may differ from the extended, star-forming galactic disks due to active galactic nuclei and higher star formation rates. The outer radius corresponds to the approximate edge of the star-forming disk of our simulated galaxy, with the HI disk extending significantly beyond this distance. The result of this process is shown in Fig.~\ref{fig:replicatingsofia}, which displays an unprocessed mock FIR total intensity map of the NUT galaxy generated with the MB11 magnetization model on the left panel and the same image after applying Gaussian smoothing, resampling, and annulus cuts on the right panel. We employ a similar method to simulate upcoming observations by the PRIMA telescope. Note that both PRIMA and HAWC+/SOFIA have similar angular resolutions (SOFIA: $13.6''$ at $154\,\mu$m; PRIMA: $9.3''$ at $100\,\mu$m). The advantage of PRIMA resides in its unprecedented sensitivity: PRIMA will be capable of measuring $5\,\mu$Jy arcsec$^{-2}$ in polarized flux at a $5\sigma$ detection level, mapping a $1$ square degree in $10$ hours of observing time \citep{Moullet2023}. PRIMA's sensitivity is $\sim$$4$ orders of magnitude better than HAWC+/SOFIA  \cite[$6\times10^{4}\,\mu$Jy arcsec$^{-2}$ in polarized flux at a $5\sigma$ detection level mapping a $1$ square degree in $10$ hours of observing time;][]{Harper2018}. PRIMA's sensitivity enables observations of the polarized dust emission from the galaxies in the Local Group ($\sim$$50$~kpc {--} $2$~Mpc) that were inaccessible by HAWC+/SOFIA. As an example, the typical total surface brightness of M33 ($840$ kpc) in the interarms is $\sim$$10^{3}\,\mu$Jy arcsec$^{-2}$ \citep{Kramer2010}. If we take the median polarization fraction of $3.3\pm0.9$\% at $154\,\,u$m across the disk in spiral galaxies measured by SALSA \citep{Lopez-Rodriguez2022b} and a $5\sigma$ detection level, then the expected polarized flux is $\sim$$7\,\mu$Jy arcsec$^{-2}$. Given that the turbulence coherence length is $\sim$$50-100$\,pc \citep{Haverkorn2008}, we use a typical distance of $\sim$$0.5$~Mpc, which provides a physical resolution of $20$~pc for an angular resolution of $9.3''$ at $100\,\mu$m by PRIMA. 

We illustrate this in Fig.~\ref{fig:front_figure}, where we show the signal-to-noise (SNR) ratio for SOFIA-like and PRIMA-like synthetic observations of the polarized intensity (PI), as observed from the galaxy in the MB12 model. The top left panel shows a standard SALSA-like observation of a local galaxy with SOFIA, resolving scales of $\sim$300~pc. The bottom left panel shows the SOFIA-like expectation for the same system, situated at a distance of 2~Mpc. Despite the higher resolution, sensitivity limitations bring most of the galaxy below the detection limit. The bottom right panel shows the same galaxy at a distance of 2~Mpc, as would be observed with a PRIMA-like instrument. Due to its significantly higher sensitivity, the ISM of the galaxy is well-resolved with high SNR. The top right panel shows the same galaxy synthetically observed in the optical $[u,g,r]$ SDSS filters\footnote{We model each stellar particle as a single stellar population \citep{Bruzual2003}. Dust absorption is modeled as an absorption screen.}, for illustration purposes. This spatial resolution resolves the gas density structures and turbulence scale in our synthetic images, which allows us 1) to study the effect of the magnetic turbulence across the galaxy's disk, 2) to provide predictions in future FIR polarimetric observations of local galaxies.

\subsection{The Alignment Angle Metric} \label{sec:alignmentanglemetric}

To quantify the alignment of the magnetic field with density structures, we first calculate the angular orientation of the magnetic field in the plane of the sky(POS),

\begin{equation}
\theta_{\mathrm{obs}} = \frac{1}{2} \arctan\!\Big(\frac{U}{Q}\Big) + \frac{\pi}{2},
\label{eq:orientation}
\end{equation}

using a four-quadrant inverse tangent. Secondly, we calculate the normalized orthogonal components of the density gradient in the image plane. The density gradient angle is oriented perpendicularly to the isocontours of total intensity, and is thus orthogonal to structures such as galactic spiral arms or dense ISM filaments.

We compute the angular separation between two orientations in the plane of the sky, $\theta_1$ and $\theta_2$, by taking the absolute value of the angular difference equation \citep{2016A&A...586A.135P}:

\begin{equation}
    \Delta\theta = \frac{1}{2}\left|\arctan{ (\frac{
    \sin{2\theta_1}\cos{2\theta_2}-\cos{2\theta_1}\sin{2\theta_2}
    }{
    \cos{2\theta_1}\cos{2\theta_2}+\sin{2\theta_1}\sin{2\theta_2}
    })}\right|.
\label{eq:ang_separation}
\end{equation}

Using Eq.~\ref{eq:ang_separation}, we define $\Delta\theta_\text{mock}$ as the angular separation where $\theta_1$ is the orientation of our mock observed B-field and $\theta_2$ is the orientation of our density gradient. An angular separation of  $\Delta\theta_\text{mock}$ = 90\degree{} between the density gradient and the magnetic field corresponds to perfect alignment between the orientation of the magnetic field lines and the isocontours of density structures. Conversely, an angular separation of 0\degree{} indicates magnetic field lines that are fully orthogonal with respect to the density structures. The angular separation between these two quantities cannot be greater than 90\degree{} when measured from polarimetric observations, because dust polarization observations trace the orientation of the magnetic field rather than the direction of the magnetic field. 

In addition, we define $\Delta\theta_\text{measurement}$ as the angular separation between the orientation of the observationally-inferred B-field, $\theta_1$, and the orientation of the intrinsic B-field, $\theta_2$, using eq.~\ref{eq:ang_separation}. Both are computed from the components of the magnetic field in the plane of the sky, $\mathbf{B}_{\text{POS}}$, as defined in Section~\ref{sec:ext_quantities}. A separation of $\Delta\theta_\text{measurement}=0^{\circ}$ between these two angles corresponds to perfect alignment, and $\Delta\theta_\text{measurement}=90^{\circ}$ to fully orthogonal measurements.

When determining the intrinsic alignment, $\Delta\theta_\text{intrinsic}$, we calculate the dot product between the normalized surface density gradient and the normalized intrinsic magnetic field vectors in the POS ($\mathbf{B}_{\text{POS}}$ defined in Section \ref{sec:ext_quantities}). We then calculate the inverse cosine of the absolute value of the dot product, which yields the angular separation between the orientation of the surface density gradient and the orientation of the intrinsic B-field,

\begin{equation}
\Delta\theta_\text{intrinsic} = \cos^{-1} \left( \left| \frac{\nabla \Sigma}{\|\nabla \Sigma\|} \cdot \frac{\mathbf{B}_{\text{POS}}}{\|\mathbf{B}_{\text{POS}}\|} \right| \right),
\end{equation}

\noindent where $\Delta\theta_\text{intrinsic}$ is the angular separation between the orientation of the surface density gradient and the orientation of the intrinsic B-field, $\Sigma$ represents the surface density, $\nabla \Sigma$ is the gradient of the surface density, $\mathbf{B}_{\text{POS}} = (B_x, B_y)$  is the intrinsic magnetic field vector in the POS, and $\|\cdot\|$ denotes the magnitude of a vector.

\subsection{Polarization fraction, angular dispersion, and magnetic alignment parameter}
\label{ss:polarization_dispersion}
To characterize how the coherence of the local magnetic field orientation affects the degree of polarization of the FIR emission, we compute three more observables, described in this section. These are the linear polarization, the polarized orientation angular dispersion, and the magnetic alignment parameter $\zeta$ \citep{Borlaff2023}, which quantifies the local alignment between the plane-of-sky magnetic field and the mean large-scale spiral pattern.

\subsubsection{Linear polarization fraction}

The mock FIR Stokes $Q$ and $U$ are observable descriptors of the FIR polarization state, from which we measure the FIR linear polarization fraction as:

\begin{equation}
    P = \frac{\sqrt{Q^2 + U^2}}{I}.
    \label{eq:pol_fraction}
\end{equation}

\subsubsection{Polarization orientation angular dispersion}
The per-pixel polarization orientation angular dispersion, S, was calculated as the circular standard deviation of all pixels within a disk around that pixel. The radius of the disk calculated over was equal to one beam FWHM of the data, so 10 pc, 13.6'' (~300 pc), and 9.3'' (~20 pc) for the simulation, SOFIA, and PRIMA, respectively. The circular standard deviation for polarization pseudovectors is defined as
\begin{equation}
\label{eq:dispersion}
    S=\frac{1}{2}\sqrt{-2 \log(\bar R)}
\end{equation}
Where $\bar R$ is the resultant length 

\begin{equation}
\label{eq:resultant}
\begin{aligned}
    \bar R = \frac{1}{N}\sqrt{\left( \sum \sin2\theta \right)^2+\left( \sum \cos2\theta \right)^2} \\
\end{aligned}
\end{equation}

$\theta$ is the angle of the polarization vector. The factor of $\frac{1}{2}$ in eq. \ref{eq:dispersion} and the $2 \theta$ in eq. \ref{eq:resultant} account for the 180$^\circ$ ambiguity in polarization pseudovectors. 

\subsubsection{Magnetic alignment parameter}

To quantify the local alignment between the plane-of-sky magnetic field and the mean large-scale spiral pattern, we compute the magnetic alignment parameter \citep{Borlaff2023},

\begin{equation}
    \zeta = \cos\!\big(2\,\Delta \theta_{\text{spiral}}),
    \label{eq:zeta}
\end{equation}

where $\Delta \theta_{\text{spiral}}$ is the difference between the local and expected polarization orientation, based on the spiral structure of the galaxy (i.e., magnetic pitch angles). To compute the expected orientation of the spiral structure for a given pixel, required for the calculation of $\zeta$, we define the tangent orientation with respect to a circle centered on the galaxy, $\theta_\text{tan}$, as:
\[
\phi = \arctan\!\Big(\frac{y - y_0}{x - x_0}\Big), \quad 
\theta_{\mathrm{tan}} = \phi + \frac{\pi}{2}.
\]
Here, $(x_0, y_0)$ denotes the galaxy center in pixel coordinates, and the inverse tangent accounts for four angular quadrants. We compute the angular separation between $\theta_1 = \theta_{\text{obs}}$ and $\theta_2 = \theta_{\text{tan}}$ in the plane of the sky (Eq.~\ref{eq:ang_separation}) to derive the pitch angle $\psi = \Delta\theta$. We can employ the pitch angle to characterize the large-scale spiral structure. We divide the angle orientation 2D map into concentric radial annuli with a fixed width of 2 pixels (Nyquist sample; $\sim10$~pc for simulation resolution, $\sim20$~pc for PRIMA resolution, and $\sim300$~pc for SOFIA resolution) and compute the circular mean pitch angle within each annulus, $\psi_{\text{mean}}$. We then compute $\Delta \theta_{\text{spiral}}$, the angular difference between the local pitch angle and the annulus mean pitch angle, via Eq.~\ref{eq:ang_separation} with $\theta_1 = \psi$ and $\theta_2 = \psi_{\text{mean}}$. 
This provides us with 2D maps of $\zeta$, where $\zeta = 1$ indicates perfect alignment of the local field with the inferred average axisymmetric spiral B-field. Conversely, $\zeta = -1$ represents perpendicularity of the local field with respect to the spiral B-field.

\subsection{Density Binning}
\label{sec:densitybinning}
\begin{figure*}
\centering
\includegraphics[width=\textwidth]{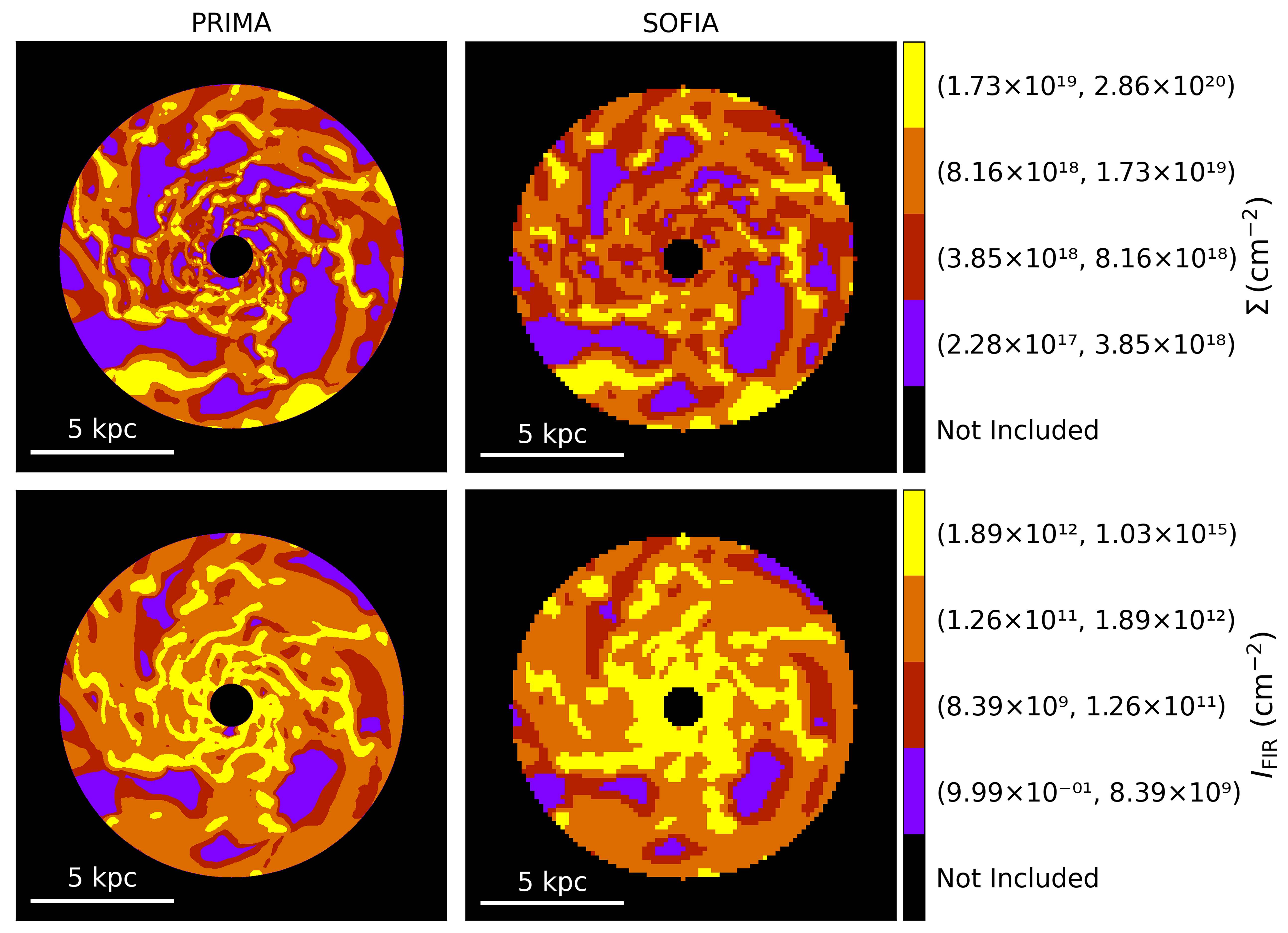}
\caption[Density Binning Map]{Regions corresponding to gas surface density and FIR total intensity bins for an individual snapshot of the MB12 model. The top row corresponds to the four density bins involving surface gas column density. The bottom row corresponds to the four total intensity bins employed to separate FIR total intensities. The left column shows the PRIMA-like case, assuming a physical resolution of $20$~pc. The right column displays instead our SOFIA-like case, which assumes a physical resolution of $300$~pc. This choice of binning corresponds to well-defined regions of the galaxy, such as the spiral arms or inter-arm regions.}
\label{fig:binfigure}
\end{figure*}

We treat the FIR total intensity as the cold-phase gas surface density for the purposes of dividing the galaxy into surface density bins. We analyze the alignment over masks that isolate specific gas surface density and FIR total intensity bins. Each gas surface density and total intensity bin corresponds to a well-defined two-dimensional region of the face-on galaxy, such as the spiral arms, inter-arm regions, or diffuse regions. The bins corresponding to different galactic regions are not all evenly spaced on a logarithmic or linear scale. We must use a quasi-asymmetrical binning method to isolate these regions across standardized bin ranges that we can compare across magnetization models. Except for the highest and lowest bin edges, the bin edges are equally spaced on a logarithmic scale. The highest and lowest bin edges are the maximum and minimum gas surface densities or FIR total intensities of all the simulation outputs for which there is a corresponding gradient and alignment measurement. Fig.~\ref{fig:binfigure} shows the gas surface density and FIR total intensity bins of the MB12 model. The highest gas surface density and FIR total intensity bins in yellow correspond to the densest parts of the spiral arms. The second-highest bins in orange correspond to more diffuse parts of the spiral arms. The third-highest bins in red correspond to the denser parts of the inter-arm regions. The lowest bins in purple correspond to the diffuse inter-arm regions. We maintain these bin ranges across all magnetization strengths and snapshots to compare the alignment properties with density for different magnetization models.

\section{Results}
\label{sec:results}

\subsection{FIR Probes of Interstellar Turbulence and Structural Alignment} 
\label{ss:magnetic_turbulence}

To shed light into the intimate interrelation between turbulence, magnetic field strength, and alignment between density structures and magnetic field orientation, we analyze the variation of magnetic turbulence across different surface density environments. While we focus in this section on addressing the turbulence in the magnetic field, we note that we find magnetic turbulence is also a close proxy for kinematic turbulence, as further discussed in Appendix~\ref{ap:turbulence}.

\begin{figure}
\centering
\includegraphics[width=\columnwidth]{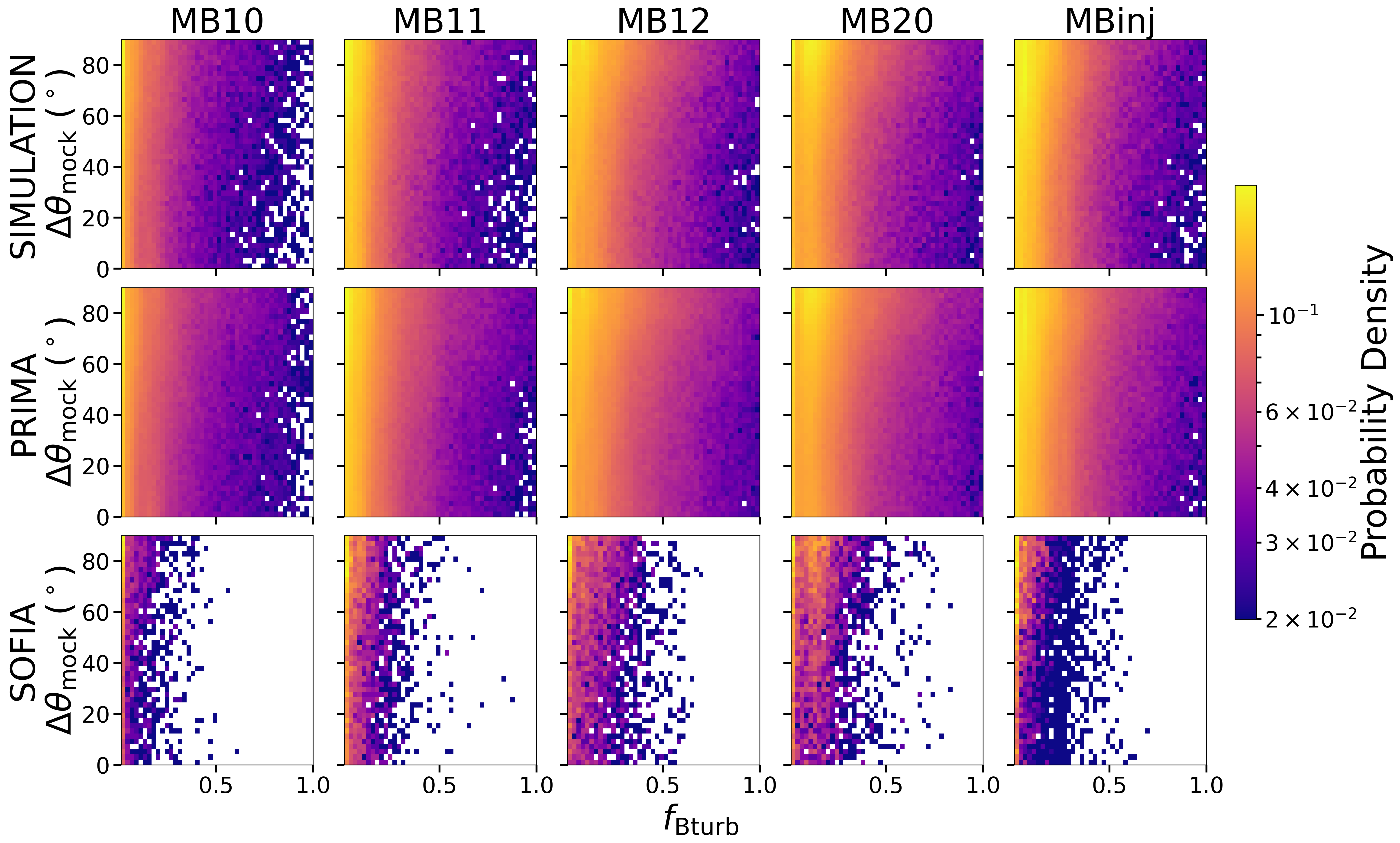}
\caption[2D Histogram of Turbulence Fraction vs Mock Observed Alignment]{Probability distribution function of our mock FIR observations across the magnetic turbulence fraction--observed alignment space. The rows correspond to the SIMULATION intrinsic configuration (top), our PRIMA-like configuration (middle), and the SOFIA-like configuration (bottom). Each column displays a different magnetization model, from the strongest (leftmost) to the weakest (fourth column). The rightmost column shows our astrophysical magnetization model.}
\label{fig:2dhist}
\end{figure}

Fig.~\ref{fig:2dhist} shows the comparison between the magnetic turbulence fraction on $100$~pc scales ($f_\text{Bturb}$) and the observed alignment {---} that is, the angular separation between the mock FIR total intensity gradient and the mock FIR-derived B-field orientation ($\Delta \theta_\text{mock}$). The 2D histograms display the distribution of the pixel values for all snapshots studied in each simulation.

To further investigate the turbulence across the distribution of magnetic field orientations, in Fig.~\ref{fig:2dhist} we show a 2D histogram across the phase space of magnetic turbulence fractions on $100$~pc scales ($f_\text{Bturb}$) and the observed alignment {---} that is, the angular separation between the mock FIR total intensity gradient and the mock FIR-derived B-field orientation ($\Delta \theta_\text{mock}$). We find a pattern of increasing $f_\text{Bturb}$ with weaker magnetization across all resolutions. Weaker magnetizations are thus correlated with higher magnetic turbulence on $100$~pc scales. We find a weak tendency for the higher turbulent fractions to take place in regions of high alignment between the CNM and the observed magnetic field. 

\begin{figure}
\centering
\includegraphics[width=0.8\columnwidth]{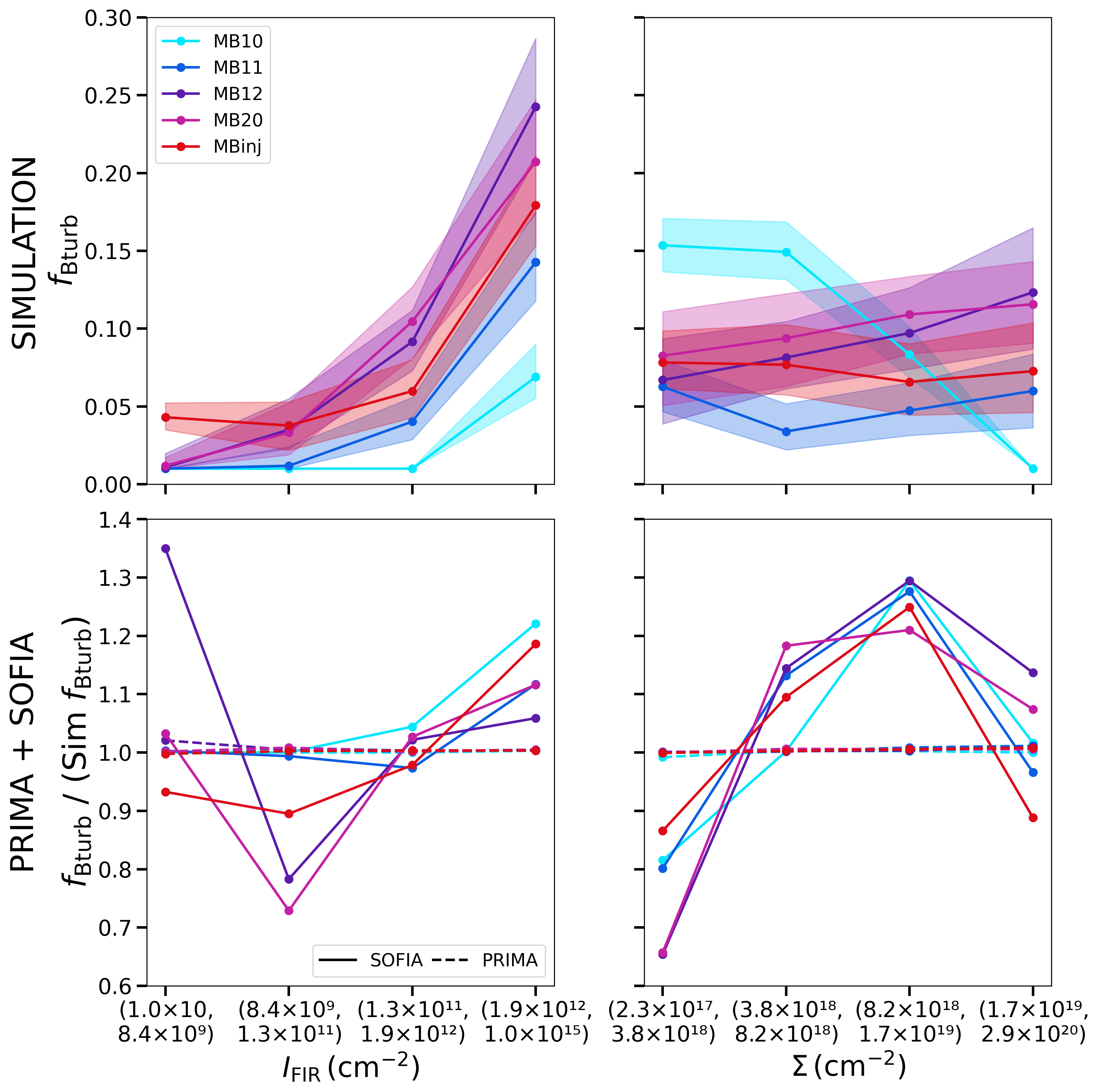}
\caption[Median Magnetic Turbulence Fraction vs. Density Range]
{Median magnetic turbulence fraction on $100$~pc scales ($f_\text{Bturb}$) and at simulation resolution, as a function of intensity ($I_\text{FIR}$; top left panel) and surface density ($\Sigma$ top right panel) ranges, as specified in Fig.~\ref{fig:binfigure}. Each color corresponds to different magnetization models. Shaded bands display the 40th and 60th quantiles. The bottom row shows the same quantities now as ratios of $f_\text{Bturb}$ between the SOFIA-like (solid lines) and PRIMA-like (dashed lines) configurations, with respect to the matching magnetization model and density range in the simulation resolution panels.}
\label{fig:turbulence mean}
\end{figure}

In Fig.~\ref{fig:turbulence mean} we show the median magnetic turbulence fraction $f_\text{Bturb}$, on $100$~pc scales, with shaded bands representing the 40th and 60th quantiles around this value. Quantile values are computed from the distribution of all pixel values across all the studied maps from all simulation snapshots. This magnetic turbulence fraction is shown as a function of FIR total intensity (left panel) and total surface gas column density (right panel). The top row displays the measurement performed at the simulation resolution, whereas the bottom row shows the ratio of each model with respect to the simulation case for the SOFIA-like (solid lines) and the PRIMA-like (dashed lines) configurations. We once again find a pattern of increasing $f_\text{Bturb}$ with weaker magnetization across all resolutions, with the exception of MB12, which displays an $f_\text{Bturb}$ that is greater than or comparable to MB20 in the highest FIR total intensity range. Weaker magnetizations are thus correlated with higher magnetic turbulence on $100$~pc scales, particularly notable in the cold-phase gas of spiral arms (i.e., highest total intensity bins). Additionally, the magnetic turbulence distributions tend to display larger standard deviation when the FIR total intensity range increases. 

We find a trend of increasing magnetic turbulence with increasing total gas column for the two weakest magnetic fields (MB20 and MB12), whereas the two models with realistic field strengths \citep[MB11 and MBinj;][]{Martin-Alvarez2024, Martin-Alvarez2023} show relatively flat trends across densities. MB10, with an extreme magnetic field strength, shows a reverse trend, which we attribute to a thicker LOS column driven primarily by the WNM and WIM disk components \citep{Martin-Alvarez2020, Martin-Alvarez2024}. We examine the same trends with respect to the FIR intensity column and find that all models show a clear increase of magnetic turbulence with increasing column, further confirming the WNM and WIM-driven thicker LOS column causing the MB10 deviation. The lower $f_\text{Bturb}$ values measured in the total gas surface density column show the higher coherence of the magnetic field orientation across a line of sight in most of the ISM. The higher values of $f_\text{Bturb}$ in the left column reflect the higher degree of magnetic turbulence in the denser media, where both dust and the FIR emission concentrate.
While the observational configurations reproduce the discussed trends with respect to increasing column densities and variations in magnetizations, the SOFIA case has deviations of up to $\sim30\%$ with respect to ground truth values from the simulation.

\subsection{The Alignment of Magnetic Fields with the Density Structures of the ISM}  
\label{ss:magnetic_alignment}

\begin{figure*}
\centering
\includegraphics[width=\textwidth]{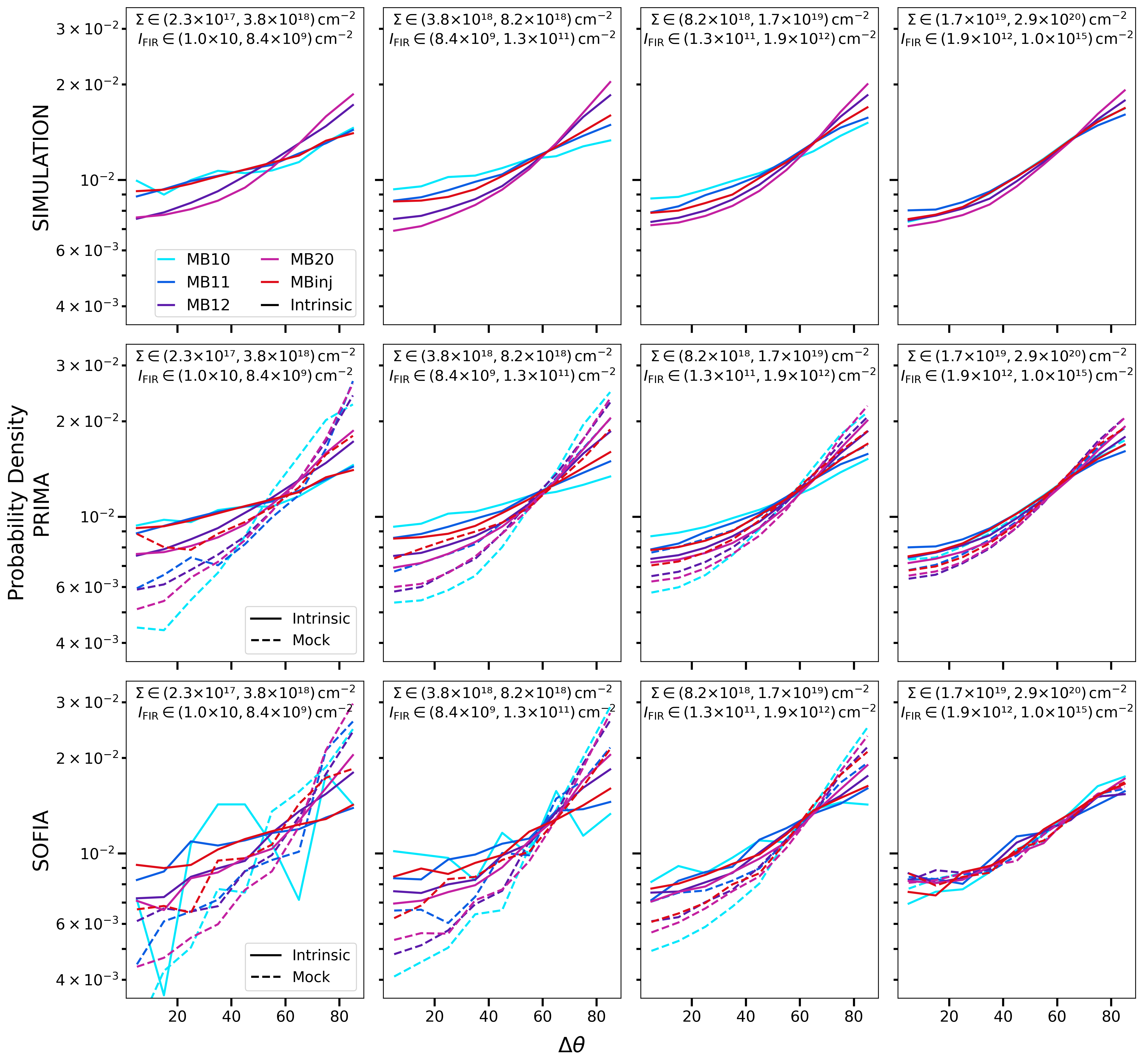}
\caption[Distributions of Intrinsic and Mock Alignment Angles, by Surface Gas Density and Total Intensity Range]{Probability density function for the intrinsic alignment ($\Delta\theta_{\rm{intrinsic}}$) angular separation between the density gradient and magnetic field as measured for the intrinsic magnetic field and total gas density surface density gradient (solid lines). We show the measured alignment between the FIR-inferred magnetic field orientation and FIR total intensity density gradient (dashed lines). Different rows correspond to different projection configurations: full resolution (top row; `SIMULATION'), PRIMA-like (middle row), and SOFIA-like (bottom row). From left to right, each column displays increasing density ranges, as correspondingly specified at the top of each panel.}
\label{fig:combinedintrinsicandmock}
\end{figure*}

We analyze the alignment between observed density structures and the inferred orientation of the magnetic field, as measured from our 2D projected maps. We show in Fig.~\ref{fig:combinedintrinsicandmock} (computed as outlined in Section \ref{sec:alignmentanglemetric}) the alignment distributions of the surface gas column density gradient with the B-field intrinsic to the simulation in the POS, as a function of gas surface density. In the first row, labeled `SIMULATION', we show the alignment PDF at the full resolution of the simulation. We find that the alignment of the magnetic field orientation with surface density structures decreases as the strength of the magnetic field increases. Higher surface densities (right) are less sensitive to variations across different magnetizations. The second row displays the same intrinsic alignment measurement (i.e., total magnetic field alignment with respect to the gas surface density structures) as a set of solid lines, now employing our PRIMA-like configuration. The resulting measurements are approximately equivalent to those performed in the `SIMULATION' row, highlighting the capabilities of PRIMA. The observational quantities are the FIR total intensity and the magnetic field orientation as inferred from the FIR Stokes $Q$ and $U$. The alignment between these two observable quantities is shown as a set of dashed lines in the same panels. Both the intrinsic and observable quantities recover the same trend. We find that PRIMA-like measurements recover smoother trends, particularly in low-surface density regions. 
These results are only weakly sensitive to minor variations in galaxy inclination, as detailed in Appendix~\ref{ap:inclination}.


\subsection{The Angular Separation between FIR-Measured and Intrinsic Magnetic Fields across Density Environments}
\label{ss:angular_measurement}

\begin{figure*}
\centering
\includegraphics[width=\textwidth]{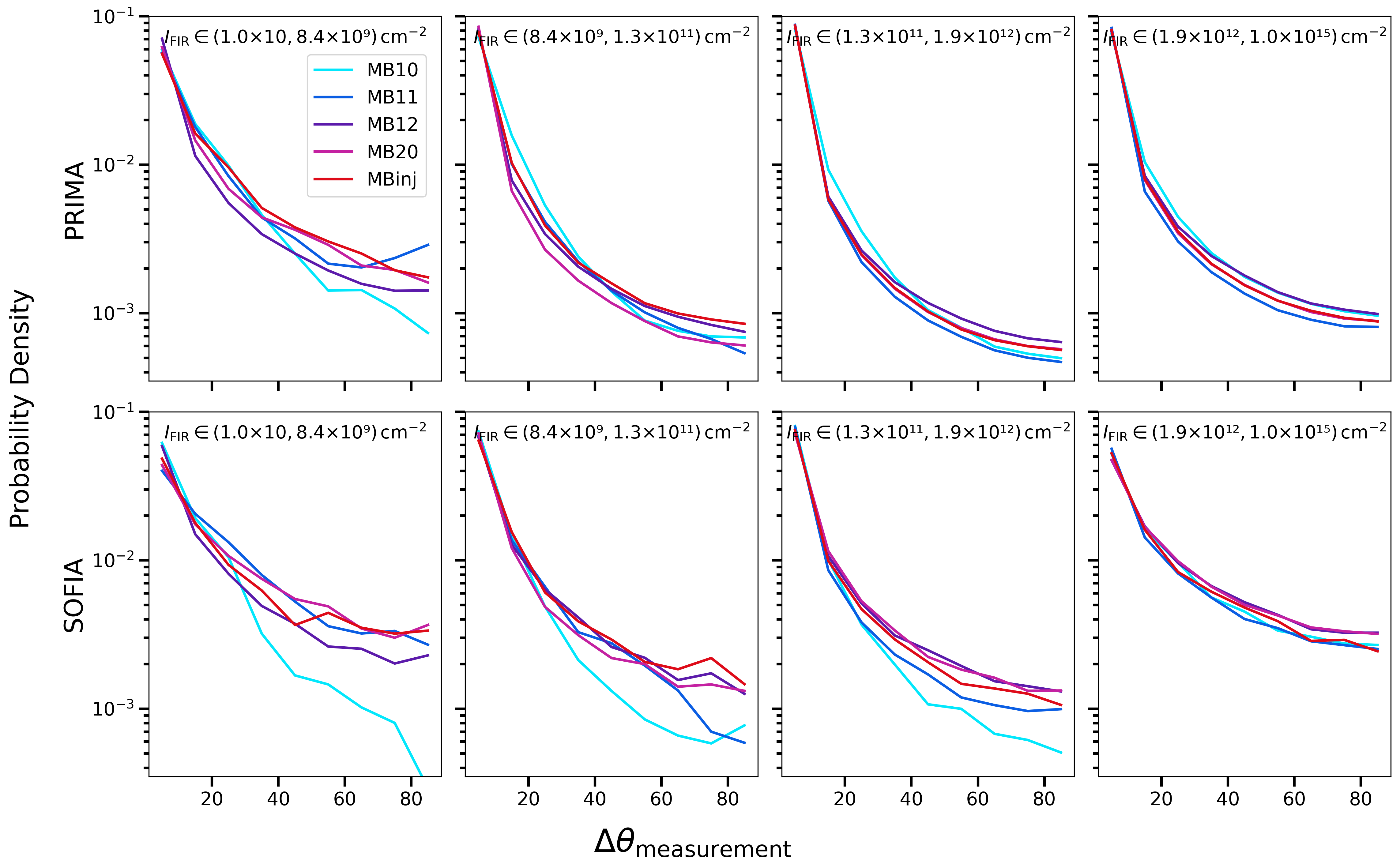}
\caption[Distributions of Angular Separations Between the Column Average B-field and the Mock FIR Derived B-field in the POS, by FIR Total Intensity Range]{Probability density distributions of angular separations between the intrinsic B-field of the simulation and the mock FIR derived B-field in the POS. Each panel contains data corresponding to specific total intensity ranges, indicated at the top of each panel. As we are directly comparing the separation between two orientations, $0^{\circ}$ angular separations correspond to alignment between both magnetic fields, whereas $90^{\circ}$ angular separations correspond to perpendicularity between both magnetic fields.
}
 \label{fig:Stokes_vs_B_by_I}
\end{figure*}

As FIR polarization traces primarily the magnetic field in the CNM phase, which dominates the ISM mass budget but occupies only a small fraction of its volume, extragalactic measurements of magnetic fields measured via FIR polarization are dominated by small, dense clumps/filaments in these systems. Depending on the resolutions accessible by observations, these magnetic field orientation measurements will be biased by beam effects. To understand how PRIMA compares with HAWC+/SOFIA, Fig.~\ref{fig:Stokes_vs_B_by_I} shows the angular separation between the magnetic field orientation as measured by our FIR observations and the intrinsic magnetic field of the simulation. The top row shows this for our PRIMA-like configuration, taking advantage of its enhanced sensitivity, whereas the bottom row shows our example of HAWC+/SOFIA. Each column corresponds to a different FIR total intensity range, increasing from left to right. Overall, we predict that both previous observations by SOFIA, and upcoming ones with PRIMA lead to measurements of the magnetic field orientation that are representative of the local magnetic field, when considered in a total gas density-weighted manner. Despite this, the precision of previous observations is somewhat sensitive to the strength of the field, and is lower for denser regions, due to these regions being notably smaller in size than the beam size. For our PRIMA-like observations we predict magnetic field orientations that remain representative of the true magnetic field orientation down to scales of $\sim20$~pc, with a $\Delta \theta \sim 6^{\circ}$ median standard deviation as measured from the PDFs of angular separations between intrinsic and measured B-field orientation.

\subsection{Observational Estimates of Magnetic Field Disorder: Polarization Fraction, Angular Dispersion, and Alignment Parameter}
\label{sec:PZetaSresults}

We investigate complementary diagnostics of magnetic field disorder, as defined in Section~\ref{ss:polarization_dispersion}. Specifically, we investigate how the polarization fraction $P$ varies as a function of the angular dispersion function $S$ and the alignment parameter $\zeta$. We focus on the MB11 model, which is representative of the models with realistic magnetic field strengths. Each of these measurements serves as an observationally accessible estimate of magnetic field coherence and small-scale structure, particularly important for observations subject to beam depolarization effects. These quantities are additional probes of the small-scale fluctuations that contribute to some of the trends explored above.

\begin{figure}
    \centering
    \includegraphics[width=\linewidth]{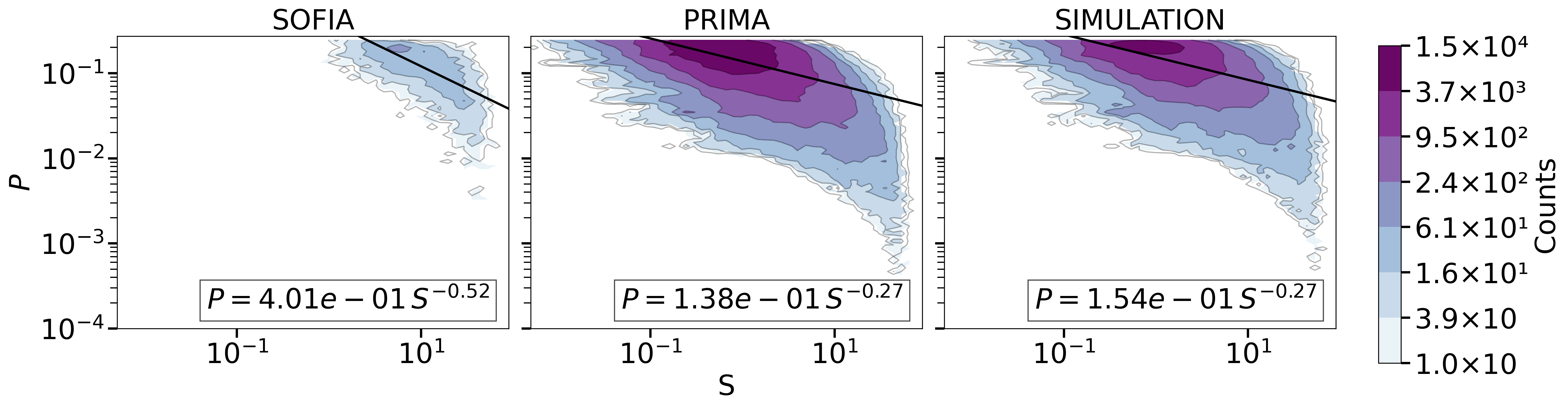}
    \caption[Per-Pixel Count Contours of the Polarization Fraction $P$ vs.\ the Angular Dispersion $S$]{Per-pixel count contours of the polarization fraction $P$ vs.\ the angular dispersion $S$, shown for the SOFIA-like observations (left), PRIMA-like observations (center) and simulation case (right). We include power-law best fit lines  in each of the panels as solid black lines.}
    \label{fig:PvsS}
\end{figure}

In Fig.~\ref{fig:PvsS}, we show the polarization fraction as a function of the angular dispersion. Panels from left to right show the SOFIA, PRIMA and simulation cases. All panels show an anti-correlation between the polarization fraction and the angular dispersion measure, with notable differences in the tightness of the correlation, and most importantly, its steepness. Measuring the power-law best-fit to the pixel distribution ($P \propto S^{\alpha}$), we find for the SOFIA case a steep depolarization ($\alpha = -0.52$) with increasing dispersion, whereas the relation is shallower for the PRIMA-like configuration, and matches that of the simulation case ($\alpha = -0.27$). Note that the P-S distribution flattens at S$<1^{\circ}$, which produces shallow slopes in the SOFIA, PRIMA, and SIMULATION distributions. This may explain the differences with the slope of $-0.834$ measured by \textit{Planck} \citep{Planck2015_overview}. Using the same range of S as \textit{Planck}, we recover a slope of $\sim-0.76$, in agreement with the \textit{Planck} observations. This shows important beam-averaging depolarization effects. Furthermore, these panels illustrate how the smaller scales accessible by PRIMA better recover intrinsic polarization fractions {--} resembling those of the simulation, which are resolved down to 10~pc. The higher resolution accessible with PRIMA will similarly enable measurements of smaller $S$ values {--} more characteristic of the diffuse ISM, where magnetic fields have higher coherence, and smaller polarization fractions {--} found in localized star forming regions which occupy a smaller volume fraction.

Vandenbroucke et al. \citep{Vandenbroucke2021} investigate the relation between these two quantities using the Auriga simulations. Specifically, they situate an observer at various locations within the Auriga Milky Way-like galaxies, and perform these measurements using the SKIRT radiative transfer code \citep{Baes2011}. This study finds an anti-correlation between these two quantities, but we note that our synthetic observations resolve higher polarization fractions and lower linear polarization angular dispersions. While some of these differences may emerge from different numerical resolutions between the simulations, magnetohydrodynamical schemes, galaxy formation models, or synthetic observational procedures, the most notable differences emerge from the observational configuration and geometries. Specifically, face-on orientations in our extragalactic observations lead to lower line-of-sight depolarization (with gas thin and thick disk scale-heights of approximately 0.2 and 1 kpc), and are mainly affected only by beam depolarization {--} leading to overall higher polarization fractions. Conversely, the resulting inferred polarization orientation on large galactic scales is relatively smooth, with its largest deviations taking place in dense clumps and star forming regions {--} with shorter coherence lengths. 

\begin{figure}
    \centering
    \includegraphics[width=0.5\linewidth]{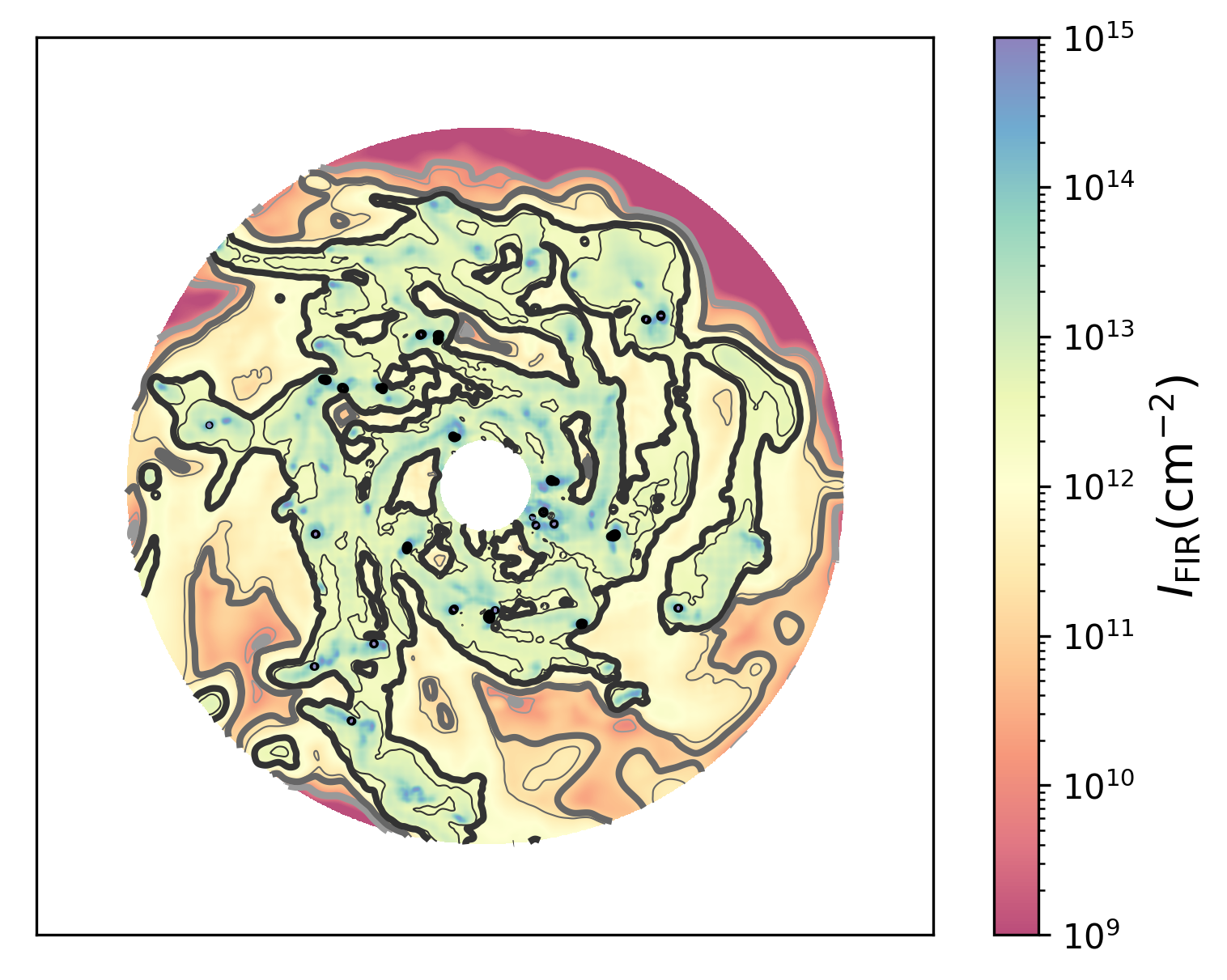}%
    \includegraphics[width=0.5\linewidth]{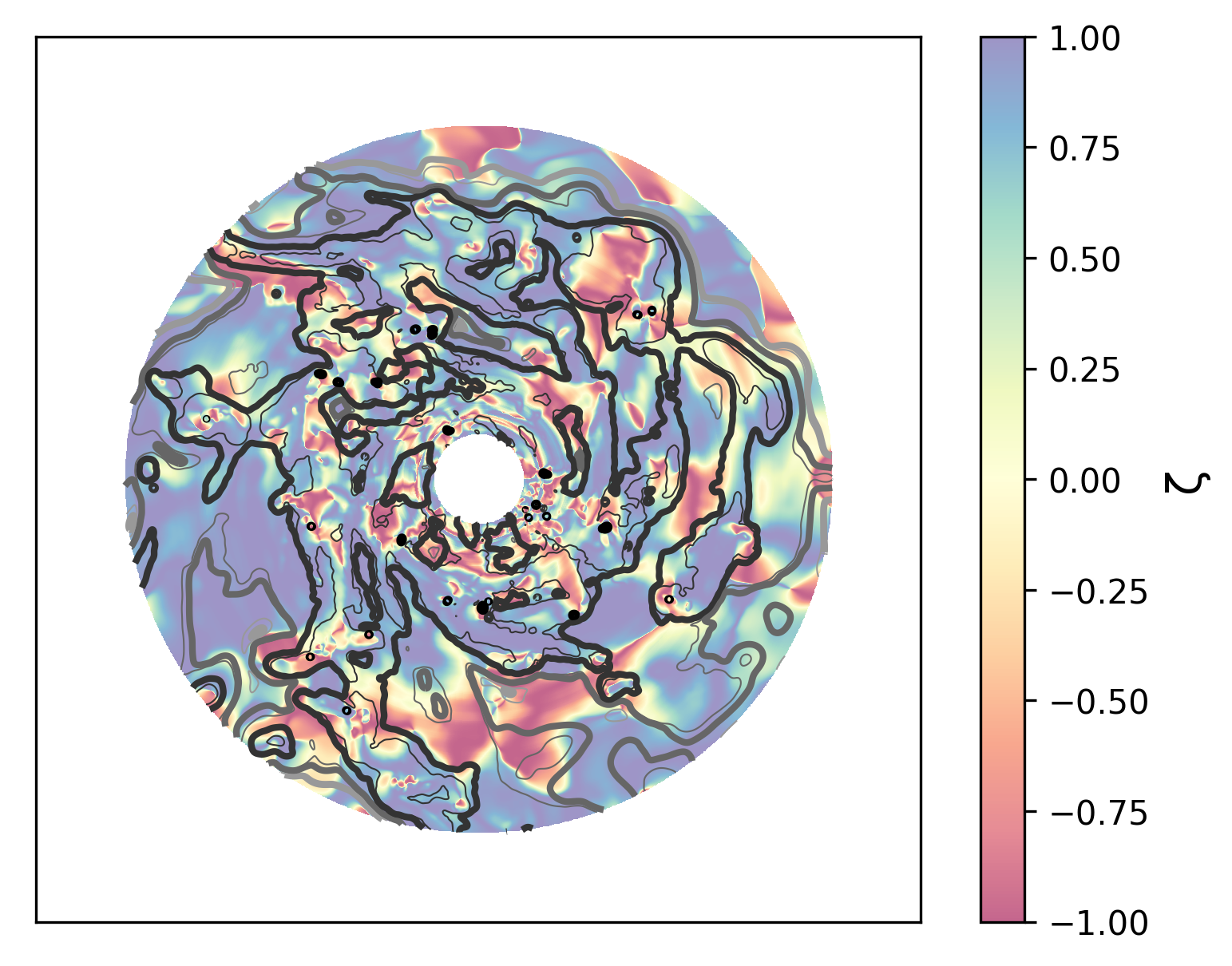}\\
    \caption[FIR Emission and Zeta Parameter Maps with Intensity Contours]{(Left panel) Synthetic FIR emission map for the MB11 model observed with PRIMA-like conditions, for guidance. Overlaid, we show thick grayscale contours marking 1~dex intervals in $I_{\mathrm{FIR}}$; darker contours encompass higher intensities. (Right panel) Map of the magnetic alignment parameter $\zeta$, quantifying the alignment between the local polarization angle and the large-scale structure of the galaxy as traced by the spiral pitch angle $\theta_\text{spiral}$. For reference, we show the same FIR contours. Regions of high $\zeta$ (blue) represent strong alignment with the  spiral magnetic field structure of the galaxy.}
    \label{fig:Zetamap}
\end{figure}

We also investigate the magnetic alignment parameter $\zeta$ \citep{Borlaff2023}, defined in eq.~\ref{eq:zeta}, which provides a simple measurement of the local alignment between the inferred magnetic field orientation and the large-scale spiral structure of the galaxy. This estimator has the advantage of being directly related to observed galaxy morphology, and reducing bias by unresolved substructures or beam-smearing effects, being well-defined and interpretable in the presence of embedded features.

An example of the resulting $\zeta$ map is shown in Fig.~\ref{fig:Zetamap}. For reference, the left panel shows the inferred FIR emission map, with overlaid contours separating 1~dex variations through thick lines. Darker lines encompass regions of higher $I_\text{FIR}$ values. The right panel of the figure displays the $\zeta$ map, with the same contours overlaid for context. Overall, we find high alignment with the spiral structure of the galaxy across most of the systems. Local deviations are found in denser regions of star formation (primarily located in density clumps within spiral arms), and in interarm regions trailing the spiral structure of the galaxy, which rotates clockwise. These results are in excellent agreement with the trends found in nearby galaxies by SALSA \citep{Borlaff2023}.

\label{ss:PvsZeta}
\begin{figure*}
\centering
\includegraphics[width=\textwidth]{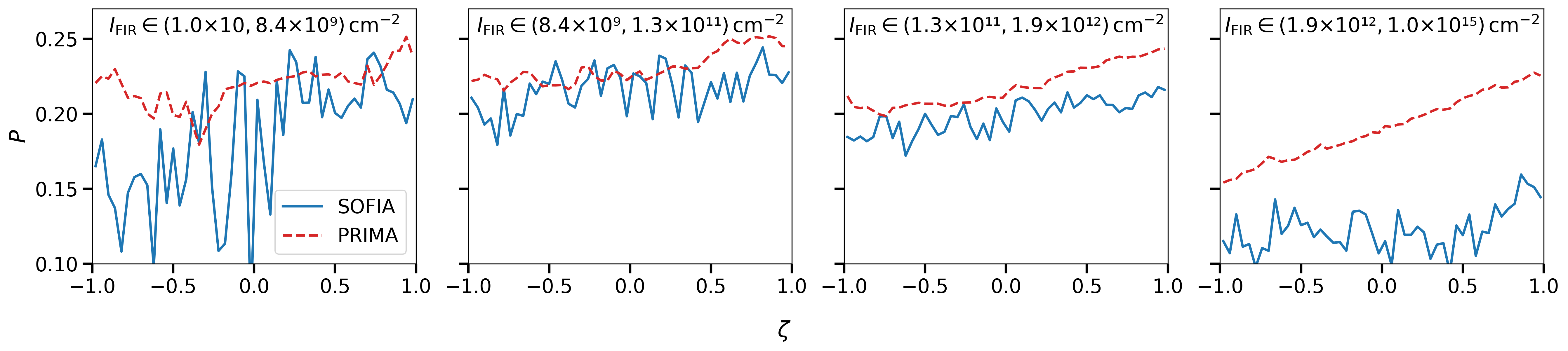}
\caption[Median Polarization Fraction $P$ vs.\ $\zeta$]{Median polarization fraction $P$ vs.\ $\zeta$ for MB11 with SOFIA-like and PRIMA-like observations at four $I_{FIR}$ ranges. Each median $P$ is calculated from values in one of $50 ~\zeta$ bins per total intensity range. In higher intensity regions, we observe that depolarization is sensitive to local deviations from the large scale magnetic field. $\zeta = 1$ represents a perfect alignment of the mean axisymmetric spiral B-field with the local field (largescale B-field dominates). $\zeta = 0$ represents perpendicularity between the mean axisymmetric spiral B-field and the local field.}
\label{fig:PvsZeta}
\end{figure*}

In Fig.~\ref{fig:PvsZeta}, we show the median $P$ as a function of $\zeta$ across our 12 snapshots of the MB11 model, separated into the four $I_\text{FIR}$ emission bins employed before. Each median $P$ is calculated from values in one of $50 ~\zeta$ bins per total intensity range. We show this relation for the SOFIA (solid blue lines) and PRIMA (red dashed lines) configurations. The panels show some positive correlation between $P$ and $\zeta$, reflecting that local alignment of the polarized emission with the spiral structure of the galaxy leads to a higher degree of polarization, as expected from the existence of large-scale coherent structures.

This positive correlation is notably less clear for the SOFIA configuration. The lowest intensity bin shows consistently low polarization fractions in a regime expected to reflect the more diffuse ISM, with higher polarizations and more coherent structures. For the two intermediate panels, we find a particularly flat relation, where the increase as a function of degree of alignment is only comparable to or less significant than the scatter. Finally, for the highest intensity range, we observe once again a relatively flat relation until $\zeta \gtrsim 0.5$.

Our prediction for the future observations by PRIMA, as depicted by the red dashed lines, shows a more optimistic scenario. The two low-intensity bins display a relatively flat relation between $\zeta$ and $P$, but for higher $P$ values. This reflects the ability of the PRIMA-like configuration to recover the intrinsic polarization fraction, with local deviations as traced by $\zeta$ corresponding to structural variations within the galaxy. As we move towards higher intensity regimes, our FIR emission estimate better traces denser structures in the galaxy, which occupy smaller volumes and with morphologies that are more decoupled from the large-scale structure of the galaxies. The two right panels (and especially the rightmost one) now display a clear positive $P \propto \zeta^{\alpha}$ correlation. We attribute this increase of $P$ with $\zeta$ to line-of-sight coherence between small-scale structures (e.g., dense clouds or spiral features) and the large-scale spiral field extending into the gas thick disk. In particularly high emission lines of sight (corresponding to high densities), more disordered magnetic field configurations and/or rapid spatial variations are expected, particularly within star-forming regions that may be subject to turbulence and feedback processes. When their alignment is high ($\zeta \to 1$), line-of-sight depolarization is minimized. The large difference we observe between the SOFIA-like and the PRIMA-like configurations provide an optimistic perspective for extending similar previous analysis \citep{Borlaff2023} to both low density regimes in distant galaxies, and high resolution observations in through the higher PRIMA sensitivity in closer ones.

\section{Conclusions} \label{sec:conclusion}

In this paper, we make use of a suite of  cosmological MHD simulations that follow the formation of a Milky Way-like galaxy to investigate the interplay between turbulence, magnetization, and the alignment of the magnetic field with density structures in the ISM. We generate various two-dimensional face-on synthetic maps that we employ to make observational predictions for the potential of the PRIMA telescope to unveil magnetic properties through extragalactic observations of local galaxies. Our main results are summarized as follows:

\begin{enumerate}
    \item We find the fraction of magnetic turbulence at 100~pc scales increases with decreasing magnetization strength. The fraction of magnetic turbulence also slightly increases as the magnetic field orientation better aligns with ISM density structures. This effect is particularly notable in the densest regions of our studied galaxy. This magnetic turbulence is also representative of kinematic turbulence (Appendix~\ref{ap:turbulence}), suggesting it may serve as an observable proxy.
    \item We find that magnetic turbulence directly correlates with the FIR intensity, which PRIMA can recover with up to $\sim4\%$ deviations from ground truth values while SOFIA can recover with up to $\sim35\%$ deviations. A similar, but less pronounced correlation is also present between magnetic turbulence and total gas surface density. Notably, extreme magnetizations (i.e. model MB10) deviate from this trend due to the impact of the magnetic field on galactic properties such as the thickening of the gas disk \citep{Martin-Alvarez2020}.
    \item We find that stronger magnetizations decrease the degree of alignment between gas density structures and intrinsic magnetic fields. 
    \item We estimate the impact of beam effects on how representative the FIR-inferred measurements of magnetic field orientation are of the global magnetic field orientation. We find that PRIMA will approximately double the precision of HAWC+/SOFIA (down to $\Delta \theta \sim 6^{\circ}$ for PRIMA and $\Delta\theta \sim 11^{\circ}$ for SOFIA), with an even higher relative precision in the densest clumps traced by the highest FIR intensity bin ($\Delta \theta \sim 8^{\circ}$ for PRIMA and $\Delta\theta \sim 19^{\circ}$ for SOFIA).
    \item SOFIA-like measurements lead to a steep anti-correlation between polarization fraction and angular dispersion ($P \propto S^\alpha$), with $\alpha = -0.52$. PRIMA-like observations are capable of recovering the shallower intrinsic $\alpha = -0.27$ also measured for the simulation. This shows the capability of PRIMA to better resolve small-scale magnetic structures, mitigating beam depolarization in regions of particularly tangled, turbulent or complex magnetic field geometries.
    \item We compare predictions for the magnetic alignment parameter $\zeta$, as introduced by the extragalactic survey SALSA with SOFIA \citep{Borlaff2023}. We find that PRIMA-like observations recover high polarization fractions in diffuse regions, where larger, coherent magnetic field structures are expected. For higher-intensity regions, where shorter coherence lengths are expected, polarization fractions are lower but display clear correlations with $\zeta$, as smaller structures better align with galactic-scale structures. This is particularly interesting when compared with SOFIA-like predictions, which have lower polarization fractions and weaker or absent correlations with $\zeta$ in high intensity bins, likely due to beam-averaging effects that obscure the underlying magnetic alignment.
\end{enumerate}

Our work highlights the unique capability of PRIMA to unveil the kinematic, magnetic, and structural properties of galaxies across multiple surface density environments probed by extragalactic observations. By shedding light on the intricate connection between different levels of magnetization and magnetic turbulence, PRIMA is well-positioned to delve into the smaller scales of galaxies, better constrain models of ISM and galactic magnetism, and expand our understanding of magnetic fields in galaxies.

\subsection*{Disclosures}
The authors have no relevant financial interests with the work presented here.

\subsection* {Code, Data, and Materials Availability} 
All employed data and simulations will be shared by the authors upon reasonable request.

\subsection*{Acknowledgments}
E.L.-R. and S.M.-A. are supported by the NASA/DLR Stratospheric Observatory for Infrared Astronomy (SOFIA) under the 08\_0012 Program. SOFIA is jointly operated by the Universities Space Research Association,Inc.(USRA), under NASA contract NNA17BF53C, and the Deutsches SOFIA Institut (DSI) under DLR contract 50OK0901 to the University of Stuttgart.
E.L.-R. and S.M.A. is supported by the NASA Astrophysics Decadal Survey Precursor Science (ADSPS) Program (NNH22ZDA001N-ADSPS) with ID 22-ADSPS22-0009 and agreement number 80NSSC23K1585. 
S.E.C. acknowledges support from NASA award 80NSSC23K0972 and an Alfred P. Sloan Research Fellowship.

\appendix

\section{The Correlation between Magnetic Turbulence and Kinematic Turbulence}
\label{ap:turbulence}

\begin{figure*}
\centering
\includegraphics[width=\textwidth]{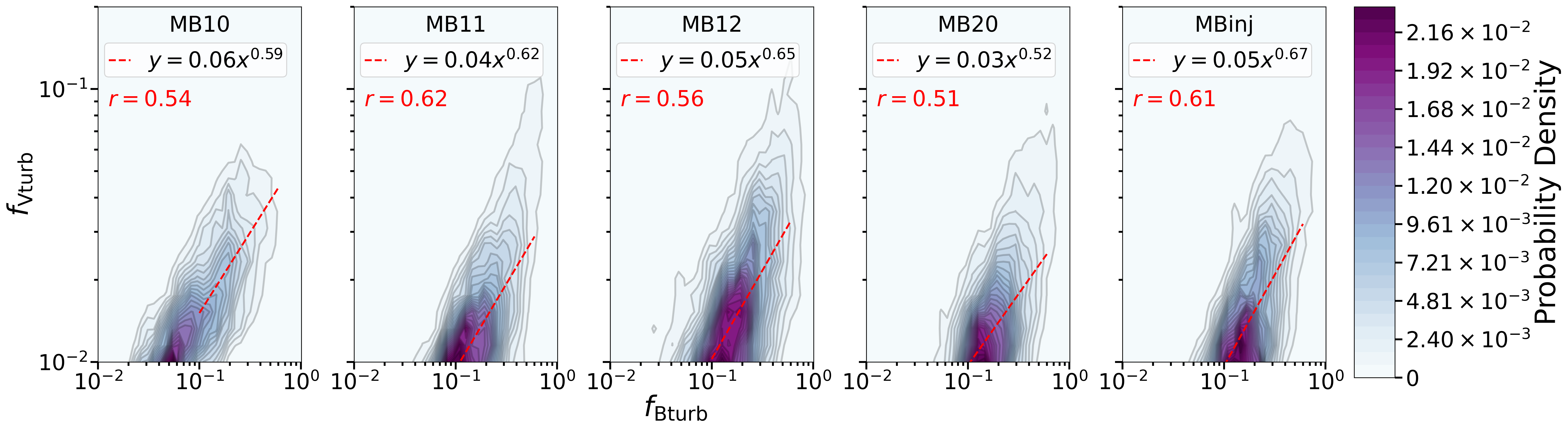}
\caption[Contour Plot of Magnetic Turbulence Fraction vs. Gas Velocity Turbulence Fraction, by Magnetization]{Contour histogram of the logarithmic scale magnetic turbulence fraction on 100~pc scales, plotted against the logarithmic scale gas velocity turbulence fraction on 100~pc scales. Power law lines of best fit are displayed with red dashed lines. The linear Pearson correlation coefficient of the logarithmically scaled data is displayed in red text. The fit only includes values $10^{-1}\leq f_{\text{Bturb}}\leq 6\times 10^{-1}$. 
}
\label{fig:contour}
\end{figure*}

Through the induction equation, magnetic fields are intimately linked to the dynamics of the ionized plasma where they reside, leading to magnetic fields frequently following bulk velocity field lines. While this suggests some degree of correlation between magnetic and kinematic turbulence, the amount of correlation between both quantities varies across different astrophysical scales and environments. To investigate this in the ISM of galaxies, we show the distribution of magnetic turbulence fraction versus the gas velocity turbulence fraction for multiple degrees of magnetization in Fig.~\ref{fig:contour}. The phase space distributions display a similar behavior for all our studied models, indicating that this direct correlation between magnetic and kinematic turbulence is preserved for all magnetizations. Despite such a high degree of correlation, we find the spread and the slope of the distribution to vary with magnetic field strength. Power law lines of best fit are calculated using linear regression in log-log space. The linear Pearson correlation coefficient of the logarithmically scaled data is displayed in red text. The fit calculation only includes values $10^{-1}\leq f_{\text{Bturb}}\leq 6\times 10^{-1}$.

\section{The Effect of Inclination on Alignment}
\label{ap:inclination}
\begin{figure*}
\centering
\includegraphics[width=\textwidth]{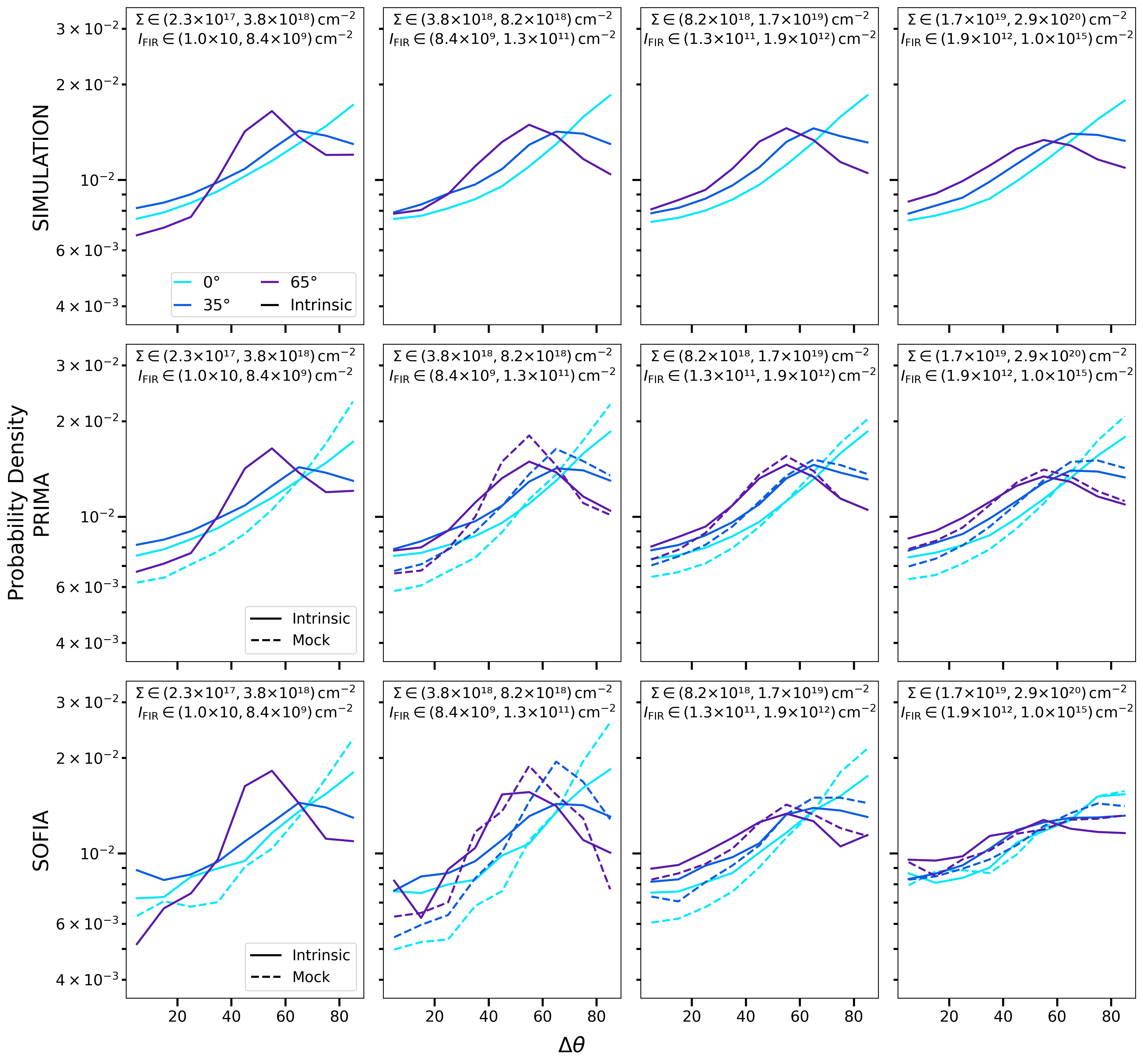}
\caption[Inclined Distributions of Intrinsic and Mock Alignment Angles, by Surface Gas Density and Total Intensity Range]{Probability distribution function for the intrinsic alignment ($\Delta\theta_{\rm{intrinsic}}$) angular separation between the density gradient and magnetic field orientation as measured for the intrinsic magnetic field and total gas density surface density gradient (solid lines). We also show the measured alignment between the FIR-inferred magnetic field orientation and FIR total intensity density gradient (dashed lines). The displayed distributions are measured in our MB12 model, with the plane of the galaxy projected at $0^{\circ}$, $35^{\circ}$, and $65^{\circ}$ inclinations with respect to the line of sight. Different rows correspond to different projection configurations: full resolution (top row; `SIMULATION'), PRIMA-like (middle row), and SOFIA-like (bottom row). From left to right, each column displays increasing density ranges, as correspondingly specified at the top of each panel. The probability density curves of alignments display a peak that grows higher and occurs at a lower alignment as the inclination angle increases.}
\label{fig:inclinedalignment}
\end{figure*}

\begin{figure*}
\centering
\includegraphics[width=\textwidth]{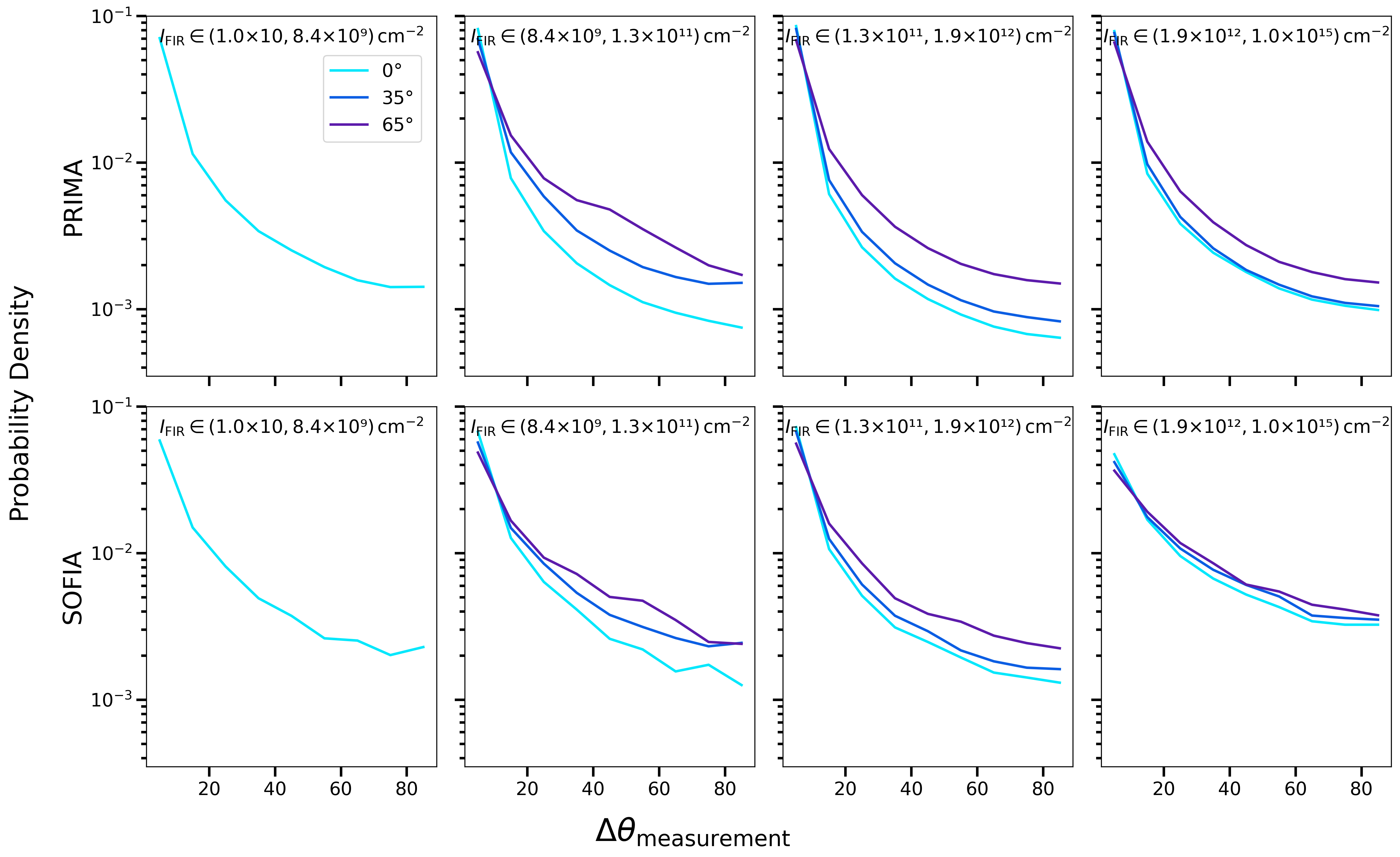}
\caption[Inclined Distributions of Angular Separations Between the Column Average B-field and the Mock FIR Derived B-field in the POS]{Probability density distributions of angular separations between the intrinsic B-field of the simulation in the POS and the mock FIR derived B-field in the POS for MB12 projected at $0^{\circ}$, $35^{\circ}$, and $65^{\circ}$ inclinations. Each panel contains data corresponding to specific total intensity($I_\text{FIR}$) ranges, indicated at the top of each panel. The smallest total intensity range in both resolution configurations only contains values from the face-on projection due to projection effects increasing the minimum column densities within the disk mask. Angular separations of $0^{\circ}$ correspond to alignment between both magnetic fields, whereas angular separations of $90^{\circ}$ correspond to perpendicularity between both magnetic fields. 
}
\label{fig:inclinedBseparation}
\end{figure*}

In Fig.~\ref{fig:inclinedalignment} and Fig.~\ref{fig:inclinedBseparation}, we display the alignment and the angular separation between the intrinsic and mock B fields under different inclinations. We conduct this analysis on all outputs of MB12 used throughout this paper, projected at $0^{\circ}$ (face-on), $35^{\circ}$ (low), and $65^{\circ}$ (high) inclinations. We follow the same procedure described in Section~\ref{sec:replicatingobservations} to generate our PRIMA-like and SOFIA-like configurations, with one key difference: the mask we use for each image is now an elliptical ``annulus". This mask corresponds to the same circular annulus described in Section~\ref{sec:replicatingobservations}, now rotated to the same inclination as the projected galaxy. 

Fig.~\ref{fig:inclinedBseparation} shows the effect on the alignment of different line-of-sight inclinations with respect to the plane of the galaxy. Due to the disk nature of the observed galaxy, orthogonal measurements of the vectors contained in this plane are now displaced to lower angular separations. For such 2D distributions and an approximate inclination measurement, this expected foreshortening can be removed from the resulting distribution through the geometric correction: 
\begin{equation}
    \theta_{\text{intrinsic}} = \arctan\left(\frac{\tan(\theta_{\text{obs}})}{\cos(i)}\right)
    \label{eq:foreshort_correction}
\end{equation}
where $\theta_{\text{intrinsic}}$ is the de-projected angle between the magnetic and intensity orientation, $\theta_{\text{obs}}$ is the observed angle between their vectors, and $i$ is the inclination angle of the disk with respect to the line of sight. The displacement of the resulting distribution shift is preserved across different telescope configurations, and the measurement is performed for the intrinsic quantities or the observationally inferred ones from the FIR estimates. This confirms that our overall results are preserved for different galaxy inclinations, particularly when considering the small scales that PRIMA will be able to resolve in nearby galaxies.

Fig.~\ref{fig:inclinedBseparation} displays a similar result to the alignment, where low observational inclinations have only minor effects on the observed distribution. At lower densities, a higher complexity along each line of sight will affect the magnetic field and the density gradients differently. While the former will be dominated by line-of-sight de-correlation, the latter will be affected by the geometries of the density structures in the ISM. We attribute the shift of angular separations to higher values we measure at $65^{\circ}$ inclination to such effects. At higher densities, we find the effects of inclination to be more negligible. When assuming the PRIMA configuration, both intrinsic and FIR-inferred quantities trace each other well. Overall, these results reinforce the validity of the proposed PRIMA analysis and observations for small observational inclinations.

\section{Changes in Zeta Distributions With Galactic Radius}
\label{ap:PSZ}

\begin{figure*}
\centering
\includegraphics[width=\linewidth]{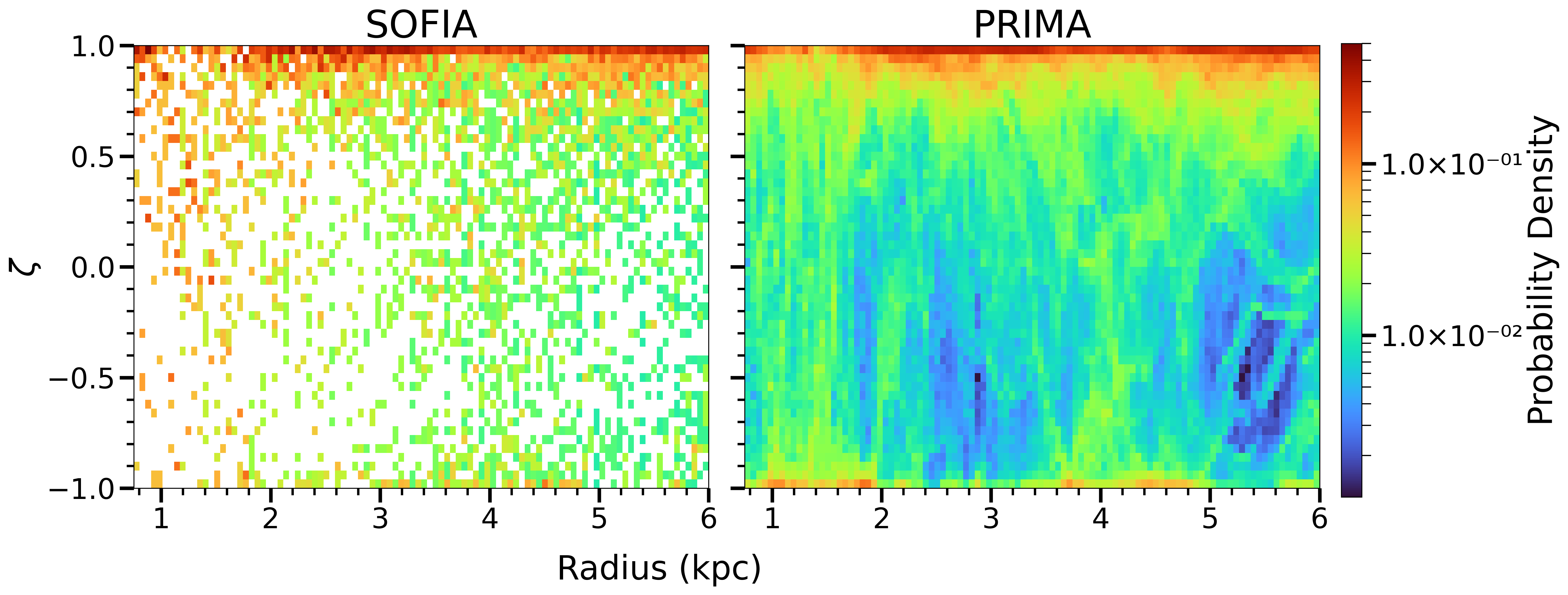}
\caption[Probability Density of $\zeta$ at Varying Radii From the Galactic Center]{Probability density of $\zeta$ at varying radii from the galactic center with SOFIA-like and PRIMA-like observations of MB11. We include 100 radial bins and 50 $\zeta$ bins, with radial bins normalized independently. $\zeta = 1$ represents a perfect alignment of the mean axisymmetric spiral B-field with the local field (largescale B-field dominates). $\zeta = 0$ represents perpendicularity between the mean axisymmetric spiral B-field and the local field.}
\label{fig:ZetavsR}
\end{figure*}

Fig.~\ref{fig:ZetavsR} shows the two–dimensional probability–density distribution of the $\zeta$ parameter as a function of galactocentric radius for the output of MB11 shown in Fig.~\ref{fig:Zetamap} when observed with SOFIA‐like and PRIMA‐like configurations. For each mock data set we divide the disc into $100$ concentric annuli and, independently, into $50$ uniform $\zeta$ bins. Within every annulus the histogram is normalised to unity, so that horizontal cuts translate directly into the relative likelihood of a given $\zeta$ value at that radius.

In both panels the highest probability density lies near
$\zeta\simeq 1$, indicating that over most radii the local magnetic field remains closely aligned with the large–scale
axisymmetric spiral component. This reflects the ordered nature of the mean field on kiloparsec scales. This result agrees with those found by \citep{Vogelsberger2014}. Conversely, the tails toward $\zeta\simeq -1$ mark sectors in which the local field is locally opposed to the mean spiral pattern, signalling either strong turbulent deflections or vertical–field excursions. A subtle radial modulation is apparent, especially in the higher–resolution PRIMA panel. Pronounced ridges with a high concentration of $\zeta\simeq 1$ occur around $R\approx2.5\text{--}3.5$~kpc and $R\approx5\text{--}6$~kpc.  These radii coincide with the locations of major spiral arms in the simulation, where large–scale shear and differential rotation reinforce the coherent spiral field. Localised pockets of negative $\zeta$ density appear near $R\approx1\text{--}2$~kpc and $3.6\text{--}4.8$~kpc. These annuli lie between arm peaks and are characterised by stronger supernova-driven turbulence, explaining their more disordered or flipped field geometry.

The SOFIA panel (left) reproduces the gross trends but is
markedly noisier and fails to recover the negative
$\zeta$ features inside $R\lesssim2$~kpc.  Beam smearing over
$\sim300$~pc averages out the small–scale reversals that PRIMA, with $\sim20$~pc resolution, captures unambiguously.  Hence PRIMA can resolve the radial oscillations in field ordering that trace the magnetised turbulence budget, while SOFIA would underestimate the degree of small‐scale disorder in the inner disc. A radial $\zeta$ profile offers a straightforward diagnostic of where and how the ordered spiral field is disrupted by turbulent processes. PRIMA‐class observations can map this diagnostic across entire galaxy discs, enabling tests of magneto-turbulent models that predict specific arm/inter-arm contrasts and radial phase shifts in field coherence.


\bibliography{main.bib}   
\bibliographystyle{spiejour}   

\listoffigures

\end{spacing}
\end{document}